\documentclass[12pt]{article}
\usepackage{natbib}
\makeatletter \def\section{\@startsection{section}{1} {\z@}{-2.5ex
plus -0.5ex minus -0.1ex}{0.5ex plus 0.1ex}{\large\bfseries}}
\def\subsection{\@startsection{subsection}{2} {\z@}{-2.25ex plus
-0.3ex minus -0.2ex}{0.05ex plus 0.05ex}{\normalsize\bfseries}}
\def\subsubsection{\@startsection{subsubsection}{3} {\z@}{-2.25ex plus
-0.3ex minus -0.2ex}{0.05ex plus 0.05ex}{\normalsize\bfseries\em}}
\makeatother
\usepackage{amsmath,amsfonts}
\usepackage{amsmath,amstext, fancyheadings}
\usepackage{amsfonts,amssymb,graphics,xspace,endnotes}
\usepackage{epsfig}
\usepackage{bm}
\usepackage{xr}
\usepackage{placeins}
\usepackage{mathptmx}

\usepackage[left=1in,right=1in,top=1in,bottom=1in]{geometry}
\numberwithin{equation}{section}

\newcommand{\ind}{{\rightarrow}_d \ }
\newcommand{\inp}{{\rightarrow}_p \ }

\newcommand{\sss}{\scriptscriptstyle}
\newtheorem{definition}{Definition}
\newtheorem{proposition}{Proposition}
\newcommand{\wpa}{\text{ wp $\rightarrow $ }}
\newcommand{\ba}{\begin{array}}
\newcommand{\ea}{\end{array}}
\newcommand{\bs}{\begin{align}\begin{split}\nonumber}
\newcommand{\bsnumber}{\begin{align}\begin{split}}
\newcommand{\es}{\end{split}\end{align}}

\renewcommand{\(}{\left(}
\renewcommand{\)}{\right)}
\renewcommand{\[}{\left[}
\renewcommand{\]}{\right]}

\newtheorem{assumption}{ASSUMPTION}
\newtheorem{theorem}{THEOREM}
\newtheorem{lemma}{LEMMA}

\begin{document}

\lhead{} \rhead{\thepage} \cfoot{} \cfoot{\fancyplain{}}
\thispagestyle{empty}
\begin{center}
\renewcommand{\thefootnote}{\fnsymbol{footnote}}
 \textbf{\Large An MCMC Approach to Classical Estimation}
\vspace{0.2in}

Victor Chernozhukov$^a$  and Han Hong$^b$

{\footnotesize \noindent $^{a}$ Department of Economics,
Massachusetts Institute of Technology, Cambridge, MA 02142, USA}
\vspace{-0.1in}

{\footnotesize \noindent $^{b}$ Department of Economics, Princeton
University, Princeton, NJ 08544, USA}

\end{center}
\begin{center}
%This Version: December 2003 \ First Version: October 2000
\end{center}
%\vspace{.1in} 
\begin{center} This is an archival version of the article "An MCMC approach to classical estimation", Journal of econometrics 115 (2), August 2003, pages 293-346.  This version does not reflect the corrections made to the article during the publication process; it  contains additional remarks and minor changes, as indicated in the text.
\end{center}

\vspace{.3in} \hrule   \noindent
\begin{center}
\textbf{Abstract}
\end{center}
\baselineskip=.88\baselineskip  {\small This paper studies
computationally and theoretically attractive estimators called the
Laplace type estimators (LTE), which include means and quantiles
of Quasi-posterior distributions defined as transformations of
general (non-likelihood-based) statistical criterion functions,
such as those in GMM, nonlinear IV, empirical likelihood, and
minimum distance methods. The approach generates an alternative to
classical extremum estimation and also falls outside the
parametric Bayesian approach. For example, it offers a new
attractive estimation method for such important semi-parametric
problems as censored and instrumental quantile, nonlinear GMM and
value-at-risk models. The LTE's are computed using Markov Chain
Monte Carlo methods, which help circumvent the computational curse
of dimensionality. A large sample theory is obtained for regular
cases.\vspace{.1in}}

 \vspace{.1in}


\baselineskip=1.1\baselineskip

 \vspace{.1in} \hrule \vspace{.2in}

\newpage

\tableofcontents

\newpage

\section{Introduction}

A variety of important econometric problems pose not only a
theoretical but a serious computational challenge, cf.
\cite{andrews:comput}. A small (and by no means exhaustive) set of
such examples include (1) Powell's censored median regression for
linear and nonlinear problems, (2) nonlinear IV estimation, e.g in
the \cite{blp1} model, (3) the instrumental quantile regression,
(4) the continuous-updating GMM estimator of \cite{cue}, and
related empirical likelihood problems. These problems represent a
formidable practical challenge as the extremum estimators are
known to be difficult to compute due to highly nonconvex criterion
functions with many local optima (but well pronounced global
optimum). Despite extensive efforts, see notably Andrews (1997),
the problem of extremum computation remains a formidable
impediment in these applications.

This paper develops  a class of estimators referred to as the
Laplace type estimators (LTE) or Quasi-Bayesian estimators
(QBE),\footnote{ A preferred terminology is taken to be the
`Laplace Type Estimators', since the term `Quasi-Bayesian
Estimators' is already used to name Bayesian procedures that use
either 'vague' or 'data-dependent' prior or multiple priors, cf.
\cite{berger:review}.} which are defined similarly to  Bayesian
estimators but use general statistical criterion functions in
place of the parametric likelihood function. This formulation
circumvents the curse of dimensionality inherent in the
computation of the classical extremum estimators by instead
focusing on  LTE which are functions of integral transformations
of the criterion functions and can be computed using Markov Chain
Monte Carlo methods (MCMC), a class of simulation techniques from
Bayesian statistics. This formulation will be shown to yield both
computable and theoretically attractive new estimators to such
important problems as (1)-(4) listed above. Although the
aforementioned applications are mostly microeconometric, the
obtained results extend to many other models, including GMM and
quasi-likelihoods in the nonlinear dynamic framework of
\cite{gallant:white}.

The class of LTE's or QBE's  aim to explore the use of the Laplace
approximation (developed by Laplace to study large sample
approximations of Bayesian estimators and for use in other
non-statistical problems) outside of the canonical Bayesian
framework -- that is, outside of parametric likelihood settings
when the likelihood function is not known. Instead, the approach
relies upon other statistical criterion functions of interest in
place of the likelihood, transforms them into proper
distributions -- Quasi-posteriors -- over a parameter of interest,
and defines various moments and quantiles of that distribution as
the point estimates and confidence intervals, respectively. It is
important to emphasize that the underlying criterion functions
are mainly motivated by the analogy principle in place of the
likelihood principle, are not the likelihoods (densities) of the
data, and are most often semi-parametric.\footnote{In this paper,
the term `semi-parametric' refers to the cases where the
parameters of interest are finite-dimensional but there are
nonparametric nuisance parameters such as unspecified
distributions.}

The resulting estimators and inference procedures possess a number
of good theoretical and computational properties and yield new,
alternative approaches for the important problems mentioned
earlier.  The estimates are as efficient as the extremum
estimates; and, in many cases, the inference procedures based on
the quantiles of the Quasi-posterior distribution yield
asymptotically valid confidence intervals, which also perform
notably well in finite samples. For example, in the quantile
regression setting, those intervals provide valid large sample
and excellent small sample inference without requiring
nonparametric estimation of the conditional density function
(needed in the standard approach). The obtained results are
general and useful -- they cover the examples listed above under
general, non-likelihood based conditions that allow
discontinuous, non-smooth semi-parametric criterion functions,
and  data generating processes that range from iid settings to
the nonlinear dynamic framework of \cite{gallant:white}. The
results thus extend the theoretical work on large sample theory
of Bayesian procedures in econometrics and statistics, e.g.
\cite{bickel2}, \cite{ibragimov}, \cite{andrews:odds},
\cite{kim:ecca}.

The LTE's are computed using MCMC, which simulates a series of
parameter draws such that the marginal distribution of the series
is (approximately) the Quasi-posterior distribution of the
parameters. The estimator is therefore a function of this series,
and may be given explicitly as the mean or a quantile of the
series, or implicitly as the minimizer of a smooth globally convex
function.

As stated above, the LTE approach is motivated by the estimation
and inference efficiency as well as computational attractiveness.
Indeed, the LTE approach is as efficient as the extremum approach,
but generally may not suffer from the computational curse of
dimensionality (through the use of MCMC). The reason is that the
computation of LTE's is itself statistically motivated. LTE's are
 typically means or quantiles
 of a quasi-posterior distribution, hence can be estimated (computed) at the
 parametric rate $1/\sqrt{B}$, where $B$ is the number of draws from that
  distribution (functional evaluations). In contrast,  the mode (extremum estimator)
 is estimated (computed) by the MCMC
and similar grid-based
 algorithms at the nonparametric rate
 $(1/B)^{\frac{p}{d+2p}}$, where $d$ is the parameter dimension and $p$ is the smoothness
 order of the objective function.

Another useful feature of LT estimation is that, by using
information about the overall shape of the objective function,
point estimates and confidence intervals may be calculated
simultaneously. It also allows incorporation of prior
information, and allows for a simple imposition of constraints in
the estimation procedure.

The remainder of the paper proceeds as follows. Section
\ref{definition} formally defines and further motivates the
Laplace type estimators with several examples, reviews the
literature, and explains other connections. The motivating
examples, which are all semi-parametric and involve no parametric
likelihoods, will justify the pursuit of a more general theory
than is currently available. Section \ref{asymptotics} develops
the large sample theory, and Sections 3 and 4 further explore it
within the context of the econometric examples mentioned earlier.
Section 4 briefly reviews important computational aspects and
illustrates the use of the estimator through simulation examples.
Section 5 contains a brief empirical example, and Section 6
concludes.

\textbf{Notation.} Standard notation is used throughout. Given
probability measure $P$, $\inp$ denotes the convergence in (outer)
probability with respect to the outer probability $P^*$; $\ind$
denotes the convergence in distribution under $P^*$, etc. See e.g.
\cite{vaart} for definitions. $|x|$ denotes the Euclidean norm
$\sqrt{x'x}$; $B_{\delta}(x)$ denotes the ball of radius $\delta$
centered at $x$. A notation table is given in the appendix.

\section{Laplacian or Quasi-Bayesian Estimation: Definition and Motivation}\label{definition}

\subsection{Motivation} Extremum estimators are usually motivated
by the analogy principle and defined as maximizers  of  random
average-like criterion functions $L_n\(\theta\)$, where $n$
denotes the sample size. $n ^{-1} L_n\(\theta\)$ are typically
viewed as transformations of sample averages that converge to
criterion functions $M\(\theta\)$ that are maximized uniquely at
some $\theta_0$.  Extremum estimators are usually consistent and
asymptotically normal, cf. \cite{amen}, \cite{gallant:white},
\cite{newey_mcfadden}, \cite{potscher}. However, in many
important cases, actually computing the extremum estimates
remains a large problem, as discussed by \cite{andrews:comput}.

\vspace{.2in} \textbf{Example 1: \textit{Censored and Nonlinear
Quantile Regression.}} A prominent model in econometrics is the
censored median regression model of \cite{powell_lad}. Powell's
censored quantile regression estimator is defined to maximize the
following nonlinear objective function \bs L_n\(\theta\) =
-\sum_{i=1}^n \omega_i\cdot \rho_{\tau} (Y_i - q(X_i, \theta)), \
\ q(X_i, \theta)=\max\(0, g(X_i,\theta\))
\end{split}\end{align}
where $\rho_{\tau}(u) = \( \tau - 1(u< 0)\) u$ is the check
function of \cite{koenker:1978}, $\omega_i$ is a weight, and $Y_i$
is either positive or zero. Its conditional quantile $q(X_i,
\theta)$ is specified as $\max(0, g(X_i,\theta))$. The censored
quantile regression model was first formulated by
\cite{powell_lad} as a way to provide valid inference in
Tobin-Amemiya models without distributional assumptions and with
heteroscedasticity of unknown form. The extremum estimator based
on the Powell's criterion function, while theoretically elegant,
has a well-known computational difficulty. The objective function
is similar to that plotted in Figure 1 -- it is nonsmooth and
highly nonconvex, with numerous local optima, posing a formidable
obstacle to the practical use of this extremum estimator; see
\cite{buchinsky}, \cite{buchinsky:phd}, \cite{fitzenberger1}, and
\cite{khan:1998} for related discussions. In this paper, we shall
explore the use of LT estimators based on Powell's criterion
function and show that this alternative is attractive both
theoretically and computationally.

\vspace{.05in}

\textbf{Example 2: \textit{Nonlinear IV and GMM.}}
\cite{amemiya:IV}, \cite{hansen:gmm}, \cite{cue} introduced
nonlinear IV and  GMM estimators that maximize \bs L_n(\theta)= -
\frac{1}{2} \(\frac{1}{\sqrt n} \sum_{i=1}^n
m_i(\theta)\)'W_n(\theta)\(\frac{1}{\sqrt n}\sum_{i=1}^n
m_i(\theta)\) +o_p(1), \end{split}\end{align}where
$m_i\(\theta\)$ is a moment function defined such that the
economic parameter of interest solves
$$
E m_i(\theta_0) =0.
$$
The weighting matrix may be given by $
W_n(\theta)=\[\frac{1}{n}\sum_{i=1}^nm_i\(\theta\)m_i\(\theta\)'\]^{-1}
+ o_p(1)$ or other sensible choices. Note that the term
``$o_p(1)$" in $L_n$ implicitly incorporates generalized
empirical likelihood estimators, which will be discussed in
section 4. Up to the first order, objective functions of
empirical likelihood estimators for $\theta$ (with the Lagrange
multiplier concentrated out) locally coincide with $L_n$.
Applications of these estimators are numerous and important (e.g.
\cite{blp1}, \cite{cue}, \cite{imbens:el}), but while global
maxima are typically well-defined, it is also typical to see many
local optima in applications. This leads to serious difficulties
with applying the extremum approach in applications where the
parameter dimension is high. As in the previous example, LTE's
provide a computable and theoretically attractive alternative to
extremum estimators. Furthermore, Quasi-posterior quantiles
provide a valid and effective way to construct confidence
intervals and explore the shape of the objective function.

\vspace{.05in}

\textbf{Example 3: \textit{Instrumental and Robust Quantile
Regression.}} Instrumental quantile regression may be defined by
maximizing a standard nonlinear IV or GMM objective
function\footnote{Early variants based on the Wald instruments go
back to \cite{mood} and \cite{hogg}, cf. \cite{koenker:1998}. }
\bs L_n(\theta)= - \frac{1}{2} \(\frac{1}{\sqrt{n}} \sum_{i=1}^n
m_i(\theta)\)'W_n(\theta)\(\frac{1}{\sqrt{n}}\sum_{i=1}^n
m_i(\theta)\),
\end{split}\end{align}
where \bs m_i\(\theta\) =   \(\tau - 1( Y_i \leq q(D_i, X_i,
\theta)\) Z_i,
\end{split}\end{align}
$Y_i$ is the dependent variable, $D_i$ is a vector of possibly
endogeneous variables, $X_i$ is a vector of regressors, $Z_i$ is
a vector of instruments, and $W_n\(\theta\)$ is a positive
definite weighting matrix, e.g.
$$
W_n(\theta)=\[\frac{1}{n}\sum_{i=1}^nm_i\(\theta\)m_i\(\theta\)'\]^{-1}
+ o_p(1) \ \text{ or } \ W_n(\theta) = \frac{1}{\tau (1-\tau)}
\[\frac{1}{n}\sum_{i=1}^n Z_iZ_i'\]^{-1},
$$
or other sensible versions. Motivations for  estimating equations
of this sort arise  from traditional separable simultaneous
equations, cf. \cite{amen}, and also more general nonseparable
simultaneous equation models and heterogeneous treatment effect
models.\footnote{See \cite{iqr} for the development of this
direction. }

Clearly, a variety of \cite{huber2} type robust estimators can be
defined in this way. For example, suppose in the absence of
endogeneity
$$
q(X,\theta) = X'\beta(\tau),
$$
then $Z= f(X)$ can be constructed to preclude the influence of
outliers in $X$ on the inference. For example, choosing $Z_{ij} =1
( X_{ij} < x_j ), \ j=1,...,\dim(X)$, where $x_j$ denotes the
median of $\{X_{ij}, i \leq n \}$ produces an approach that is
similar in spirit to the maximal regression depth estimator of
\cite{regdepth}, whose computational difficulty is well known, as
discussed in \cite{depreg:comput}. The resulting objective
function $L_n(\beta)$ is highly robust to both outliers in
$X_{ij}$ and $Y_i$. In fact, it appears that the breakdown
properties of this objective function are similar to those of the
objective function of \cite{regdepth}.

Despite a clear appeal, the computational problem is daunting.
The function $L_n$ is highly non-convex, almost everywhere flat,
and has numerous discontinuities and local
optima.\footnote{\cite{macurdy_timmins} propose to smooth out the
edges using kernels, however this does not eliminate
non-convexities and local optima; see also \cite{abadie:1995}.}
(Note that the global optimum is well pronounced.) Figure 1
illustrates the situation. Again, in this case the LTE approach
will yield a computable and theoretically attractive alternative
to the extremum-based estimation and inference.\footnote{Another
computationally attractive approach, based on an  extension of
\cite{koenker:1978} quantile regression estimator to instrumental
problems like these, is given in Chernozhukov and Hansen(2001). }
Furthermore, we will show that the Quasi-posterior confidence
intervals provide a valid and effective way to construct
confidence intervals for parameters and their smooth functions
without non-parametric estimation of the conditional density
function evaluated at quantiles (needed in standard
approach). $\square$\\

\begin{figure}[htbp!]
 \begin{center}
\noindent
 \includegraphics[width=6in,height=4in]{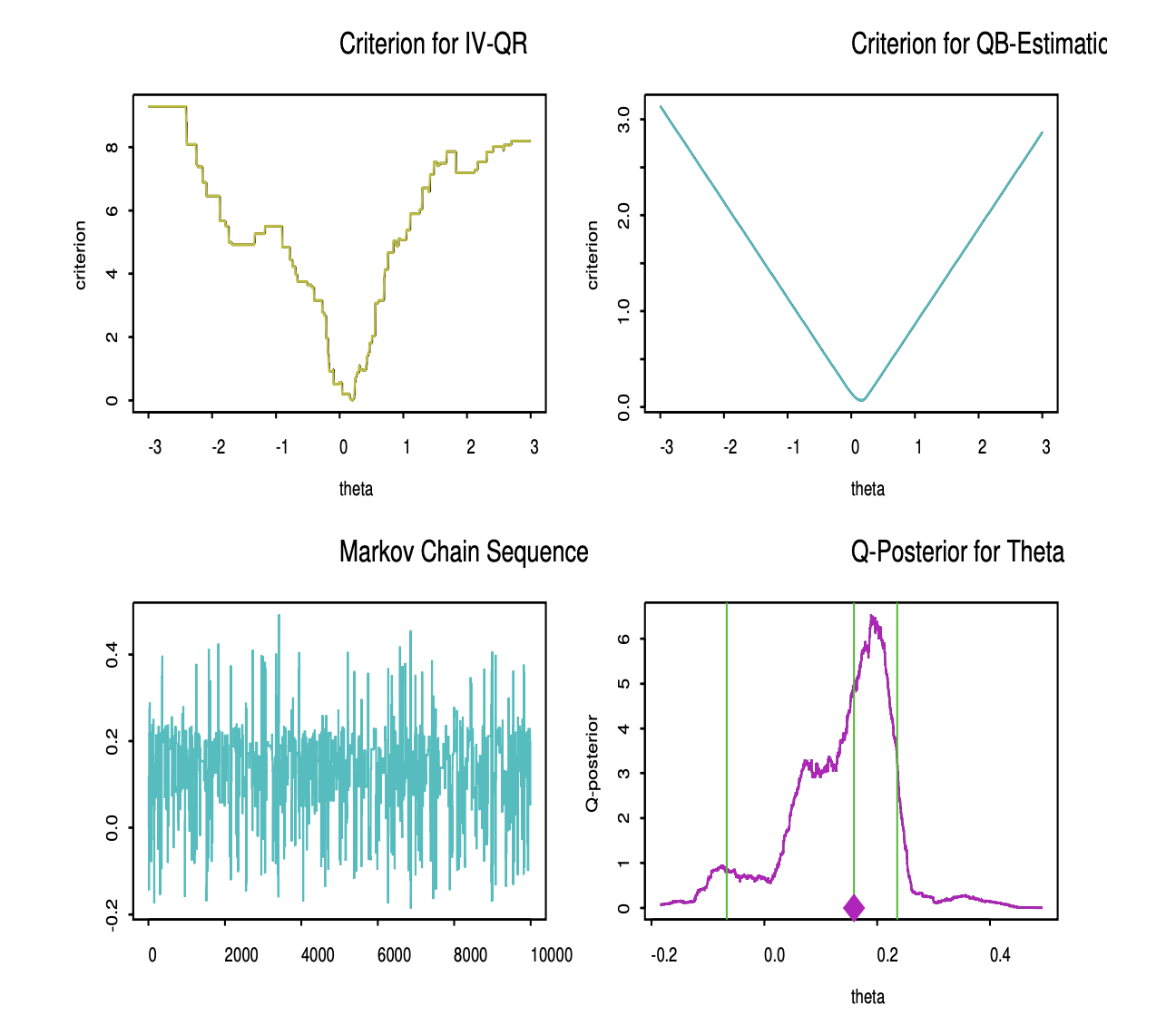}
 \caption{A Nonlinear IV Example involving
 Instrumental Quantile Regression.
 In the top-left panel the discontinuous
 objective function $L_n(\theta)$ is depicted (one-dimensional case).
 The true parameter
 $\theta_0=0$. In the bottom-left panel,
 a Markov Chain sequence of draws $\(\theta^{(1)},...\theta^{(J)}\)$
 is depicted. The marginal distribution
 of this sequence is
 $p_n(\theta) = e^{\sss L_n(\theta)}/\int_{\Theta} e^{\sss L_n(\theta)}d \theta $
 , see the bottom-right panel.
 The point estimate, the sample mean $\bar
 \theta$, is given by the vertical line with the romboid root. Two other vertical lines are the 10-th and the 90-th percentiles of quasi-posterior
 distribution. The upper-right panel depicts the expected loss function
 that the LTE minimize. }\label{Figure 1}
 \end{center}
\end{figure}

The LTE's studied in this paper can be easily computed through
Markov Chain Monte Carlo and other posterior simulation methods.
To describe these estimators, note that although the objective
function $L_n\(\theta\)$ is generally not a log-likelihood
function, the transformation \bsnumber\label{belief}
p_n\(\theta\) =
    \frac{e^{L_{\sss n}(\theta)} \pi(\theta)}
    {\int_{\Theta }e^{L_{\sss n}(\theta)}\pi(\theta) d \theta}
\end{split}\end{align}
is a proper distribution density over the parameter of interest,
called  here the \textit{Quasi-posterior}.  Here $\pi\(\theta\)$
is a weight or prior probability density that is strictly positive
and continuous over $\Theta$, for example, it can be constant over
the parameter space. Note that $p_n$ is generally not
 a true posterior in the Bayesian sense, since it
may not involve the conditional data density or likelihood, and
is thus generally created through non-Bayesian statistical
learning.

The Quasi-posterior mean is then defined as \bsnumber\label{mean}
\hat\theta  & = \int_\Theta \theta  p_n(\theta) d \theta =
\int_\Theta \theta
 \(\frac{e^{L_{\sss n}(\theta)} \pi(\theta)}
    {\int_{\Theta }e^{L_{\sss n}(\theta)}\pi(\theta) d \theta}\) d
    \theta ,
\end{split}\end{align}
where $\Theta$ is the parameter space. Other quantities such as
medians and quantiles will also be considered. A formal
definition of LTE's  is given in Definition \ref{def1}.

In order to compute these estimators, using Markov Chain Monte
Carlo methods, we can draw a Markov chain (see Figure 1), $$S =
\(\theta^{(1)}, \theta^{(2)}, ..., \theta^{({\sss B})} \),$$
whose marginal density is approximately given by $p_n(\theta)$,
the Quasi-posterior distribution.  Then the estimate $\widehat
\theta$, e.g. the Quasi-posterior mean, is computed as \bs
\widehat \theta =  \frac{1}{B} \sum_{i=1}^B \theta^{(i)}.
\end{split}\end{align}
Analogously, for a given continuously differentiable function $g:
\Theta \to \Bbb{R}$, the $90\%$-confidence intervals are
constructed simply by taking the $.05$-th and $.95$-th quantiles
of the sequence
$$
g(S) = \(g(\theta^{(1)}),  ..., g(\theta^{({\sss B})}) \),
$$
see Figure \ref{Figure 1}. Under the information equality
restrictions discussed later, such confidence regions are
asymptotically valid. Under other conditions, it is possible to
use other Quasi-posterior quantities such as the
variance-covariance matrix of the series $S$ to define
asymptotically valid confidence regions, see Section 3.  It shall
be emphasized repeatedly that the validity of this approach does
not depend on the likelihood formulation.

\subsection{Formal Definitions}

Let $\rho_n\(u\)$ be a  \textit{penalty or loss function}
associated with making an incorrect decision. Examples of
$\rho_n\(u\)$ include
\begin{itemize}
\item[ \textbf{i.} ]$\rho_n\(u\) = \vert \sqrt{n} u\vert^2$, \ \ the squared loss function,
\item[ \textbf{ii.} ] $\rho_n\(u\) = \sqrt{n}\sum_{j=1}^d \vert
u_j\vert$, \ \ the absolute deviation loss function, \item[
\textbf{iii.} ]$\rho_n\(u\) = \sqrt{n} \sum_{j=1}^d
\(\tau_j-1\(u_j\leq 0\)\)u_j$,  for $\tau_j \in (0,1)$ for each
$j$, the check loss function of \cite{koenker:1978}.
\end{itemize}

The parameter is assumed to belong to the subset $\Theta$ of
Euclidean space. Using the Quasi-posterior $p_n$ density in
(\ref{belief}),  define the Quasi-posterior risk function as:
\bsnumber\label{average_risk_2} Q_n\(\zeta\) &= \int_\Theta
\rho_n\(\theta- \zeta\) p_n\(\theta\) d\theta = \int_\Theta
\rho_n\(\theta- \zeta\) \(
    \frac{e^{L_{\sss n}(\theta)} \pi(\theta)}
    {\int_{\Theta }e^{L_{\sss n}(\theta)}\pi(\theta) d \theta} \) d
    \theta.
\end{split}\end{align}

\begin{definition}\label{def1} The class of LTE
minimize the function $Q_n(\zeta)$ in (\ref{average_risk_2}) for
various choices of $\rho_n$:
\bsnumber\label{definition_Quasi-Bayes} \hat\theta &=
\arg\inf_{\zeta\in\Theta} \Big [ Q_n(\zeta) \Big ].
\end{split}\end{align}
\end{definition}

The estimator $\widehat \theta $ is a decision rule that is least
unfavorable given the statistical (non-likelihood) information
provided by the probability measure $p_n$, using the loss
function $\rho_n$. In particular, the loss function $\rho_n$ may
asymmetrically penalize  deviations from the truth, and $\pi$ may
give differential weights to different values of $\theta$. The
solutions to the problem (\ref{definition_Quasi-Bayes}) for loss
functions \textbf{i-iii} include the Quasi-posterior means,
 medians, and  marginal $\tau_j$-th quantiles, respectively.\footnote{ This formulation implies that conditional on the data,
the decision $\widehat \theta$ satisfies Savage's axioms of choice
under uncertainty with subjective probabilities given by $p_n$
(these include the usual asymmetry and negative transitivity of
strict preference relationship, independence, and some other
standard axioms).}

\subsection{Related Literature} Our analysis will rely heavily
on the previous work on Bayesian estimators in the likelihood
setting. The initial large sample work on Bayesian estimators was
done by Laplace (see \cite{stigler:laplace} for a detailed
review). Further early work of \cite{bernstein1} and
\cite{vonmises1} has been considerably extended in both
econometric and statistical research, cf. \cite{ibragimov},
\cite{bickel2}, \cite{andrews:odds}, \cite{phillips:ploberger},
and \cite{kim:ecca}, among others.

In general, Bayesian asymptotics require very delicate control of
the tail of the posterior distribution and were developed in
useful generality much later than the asymptotics of extremum
estimators.  The treatments of \cite{bickel2} and \cite{ibragimov}
are most useful for the present setting, but are inevitably tied
down to the likelihood setting. For example, the latter treatment
relies heavily on Hellinger bounds that are firmly rooted in the
objective function being a likelihood of iid data. However, the
general flavor of the approach is suited for the present purposes.
The treatment of \cite{bickel2} can be easily extended to smooth,
possibly incorrect iid likelihoods,\footnote{See
\cite{pseudoBayes} for an extension to the more than three times
differentiable smooth misspecified iid likelihood case. The
conditions do not apply to GMM or even Example 1. } but does not
apply to censored median regression or any of the GMM type
settings. \cite{andrews:odds} and \cite{phillips:ploberger} study
the large sample approximation of posteriors and posterior odds
ratio tests in relation to the classical Wald tests in the context
of smooth, correctly specified likelihoods. \cite{kim:ecca}
derives the limit behavior of posteriors in likelihood models over
shrinking neighborhood systems. Kim's approach and related
approaches have been important in describing the essence of
posterior behavior, but the
 limit behavior of point estimates like ours does not
follow from it.\footnote{E.g. the approach does not characterize the behavior of the
posterior mean: $\int_{-\infty}^{\infty} \theta p_n(\theta) d
\theta$, which also requires the study of the complete
$L_n(\theta)$ and the analysis of convergence of the posterior in the total variation of moments norm.}

Formally and substantively, none of the above treatments apply to
our motivating examples and the estimators given in Definition 1.
These examples do not involve likelihoods, deal mostly with GMM
type objective functions, and often involve discontinuous and
non-smooth criterion functions to which the above mentioned
results do not apply.  In order to develop the theory of LTE's
for such examples, we extend the previous arguments. The results
obtained here enable the use of Bayesian tools outside of the
Bayesian framework -- covering models with non-likelihood-based
criterion functions, such as examples listed earlier and other
semi-parametric objective functions that may, for example, depend
on preliminary estimates of infinite-dimensional nuisance
parameters. Moreover, our results apply to general forms of data
generating processes -- from the cross-sectional framework of
\cite{amen} to the nonlinear dynamic framework of
\cite{gallant:white} and \cite{potscher}.

Our motivating problems are all semi-parametric, and there are
several pure Bayesian approaches to such problems, see notably
 \cite{doksum}, \cite{freedman},
\cite{hahn:bb}, \cite{imbens:chamberlain}, \cite{kottas}.
Semi-parametric models have some parametric  and nonparametric
components, e.g. the unspecified nonparametric distribution of
data in Examples 1-3. The mentioned papers proceed with the pure
Bayesian approach to such problems, which involves Bayesian
learning about these two components via a two-step process. In the
first step, Bayesian non-parametric learning with Dirichlet priors
is used to form beliefs about the joint non-parametric density of
data, and then draws of the non-parametric density (``Bayesian
bootstrap") are made repeatedly to compute the extremum parameter
of interest. This approach is purely Bayesian, as it fully
conforms to the Bayes learning model. It is clear that this
approach is generally quite different from LTE's or QBE's studied
in this paper, and in applications, it still requires numerous
re-computations of the extremum estimates in order to construct
the posterior distribution over the parameter of interest. In
sharp contrast, the LT estimation takes a ``shortcut" by
essentially using the common criterion functions as posteriors,
and thus entirely avoids both the estimation of the nonparametric
distribution of the data and the repeated computation of extremum
estimates.

Finally note that the LTE approach has a limited-information or
semi-parametric nature in the sense that we do not know or are
not willing to specify the complete data density. The
limited-information principle is powerfully elaborated in the
recent work of \cite{zellner:maxent}, who starts with a set of
moment conditions, calculates the maximum entropy densities
consistent with the moment equations, and uses those as formal
(misspecified) likelihoods. While in the present framework,
calculation of the maximum entropy densities is not needed, the
large sample theory obtained here does cover Zellner's
(\citeyear{zellner:maxent}) estimators as one fundamental case.
Related work by \cite{kim2} derives a limited information likelihood interpretation for certain smooth GMM
settings.\footnote{\cite{kim2} also provided some useful
asymptotic results for $\exp(L_n(\theta) )$ using the shrinking
neighborhood approach. However, Kim's (\citeyear{kim2}) approach
does not cover the estimators and procedures considered here, see
previous footnote. } In addition,  the LTE's based on the
empirical likelihood are introduced in Section 4 and motivated
there as respecting the limited information principle.

\section{Large Sample Properties}\label{asymptotics}

This section shows that under general regularity conditions the
Quasi-posterior distribution concentrates at the speed
$1/\sqrt{n}$ around the true parameter $\theta_0$ as measured by
the ``total variation of moments" norm (and total variation norm
as a special case), that the LT estimators are consistent and
asymptotically normal, and that Quasi-posterior quantiles and
other relevant quantities provide asymptotically valid confidence
intervals.

\subsection{Assumptions} We begin by stating the main assumptions.
In addition, it is assumed without further notice that the
criterion functions $L_n(\theta)$ and other primitive objects
have the standard measurability properties.  For example, given
the underlying probability space $(\Omega, \mathcal{F}, P)$, for
any $\omega \in \Omega$, $L_n(\theta)$ is a measurable function of
$\theta$, and for any $\theta \in \Theta$, $L_n(\theta)$ is a
random variable, that is a measurable function of $\omega$.

\begin{assumption}[\textbf{\textit{Parameter}}]\label{parameter} The true parameter
$\theta_0$ belongs to the interior of a compact convex subset
$\Theta$  of  Euclidean space $\Bbb{R}^d$.
\end{assumption}

\begin{assumption}[\textbf{\textit{Penalty Function}}]\label{loss function} The loss
function $\rho_n: \Bbb{R}^d \rightarrow \Bbb{R}_+$ satisfies:
\begin{itemize}

\item[\textbf{i.}] $\rho_n(u)=\rho( \sqrt{n}  u )$,
where $\rho(u) \geq 0$ and $\rho(u) = 0$ iff $u=0$,

\item[\textbf{ii.}] $\rho$ is convex and  $\rho(h) \leq 1 + | h
|^p$ for some $p \geq 1$,

\item[\textbf{iii.}]  $\varphi(\xi) = \int_{\Bbb{R}^d} \rho(u-\xi)
e^{\sss -u'a u} du$ is minimized uniquely at some $\xi^* \in
\Bbb{R}^d$ for any finite $a >0$,

\item[\textbf{iv.}]  the weighting function $\pi: \Theta \rightarrow \Bbb{R}_+$
is a continuous, uniformly positive density function.

 \end{itemize}
\end{assumption}

\begin{assumption}[\textbf{\textit{Identifiability}}]\label{consistency assumption}
For any $\delta > 0$, there exists $\epsilon > 0$, such that \bs
\liminf_{n \rightarrow \infty }P_*\left \{\sup_{\vert \theta -
\theta_0\vert \geq \delta} \frac1n \Big (L_n\(\theta\) -
L_n\(\theta_0\)\Big ) \leq -\epsilon \right \} = 1.
\end{split}\end{align}
\end{assumption}

\begin{assumption}[\textit{\textbf{Expansion}}]\label{second order expansion}
For $\theta$ in an open neighborhood of $\theta_0$,
\begin{itemize}
\item[\textbf{i.}] $L_n\(\theta\) - L_n\(\theta_0\) =
\(\theta-\theta_0\)'\Delta_n\(\theta_0\) -\frac12 \(\theta -
\theta_0\)' \[n J_n\(\theta_0\)\]\(\theta - \theta_0\) +
R_n\(\theta\)$,
\item[\textbf{ii.}] $\Omega_n^{\sss -1/2}(\theta_0) \Delta_n\(\theta_0\)/\sqrt{n} \
\ind \mathcal{N}\(0, I\)$,
 \item[\textbf{iii.}] $J_n\(\theta_0\)=O(1)$ and $\Omega_n(\theta_0)=O(1)$
 are uniformly in $n$ positive-definite constant matrices,
\item[\textbf{iv.}] for each $\epsilon > 0$ there is a sufficiently
small $\delta >0$ and large $M>0$ such that \bs
 \textrm{\textbf{(a)}} \ \ \limsup_{n \to \infty} \ &P^* \left \{ \sup_{  M/\sqrt{n}\leq |\theta -
 \theta_0 | \leq \delta } \frac{ | R_n\(\theta\) |} { n | \theta -
 \theta_0 |^2}> \epsilon \right \} < \epsilon, \\
\textrm{\textbf{(b)}} \ \ \limsup_{n \to \infty} \ &P^* \left \{
\sup_{ |\theta - \theta_0 | \leq M/\sqrt{n}}\ \ \ \  |
R_n\(\theta\) | \
> \epsilon  \ \right \}  =
 0.
\end{split}\end{align}

\end{itemize}

\end{assumption}

\subsection{Discussion of Assumptions} In the following we discuss
the stated assumptions under which Theorem 1-4 stated below will
be true. We argue that these assumptions are simple but encompass
a wide variety of econometric models -- from cross-sectional
models to nonlinear dynamic models. This means that Theorems 1-4
are of wide interest and applicability.

In general, Assumptions 1-4 are related to but different from
those in \cite{bickel2} and \cite{ibragimov}.
 The most substantial differences appear in Assumption 4, and are
 due to the general non-likelihood setting. Also in Assumption 4
 we introduce Huber type conditions to handle the tail behavior of discontinuous and non-smooth criterion
functions. In general, the early approaches are inevitably tied
to the iid likelihood formulation, which is not suited for the
present purposes.

 The compactness Assumption \ref{parameter}
 is conventional. It is shown in the proof of Theorem 1 that it
 is not difficult to drop compactness. For example, in the case of
Quasi-posterior quantiles in Theorem 3, it is only required that
$\pi$ is a
 proper density; in the case of
 Quasi-posterior variances in Theorem 4, it is only required that $\int_{\Theta}|\theta|^2 \pi(\theta)
 d\theta < \infty$; and for the general loss functions considered in Theorem 2 it is required that $\int_{\Theta}|\theta|^p \pi(\theta)
 d\theta < \infty$. Of course, compactness guarantees all of the above given Assumption 2.  Also, that
the parameter is on the interior of the parameter space rules out
some non-regular cases; see for example \cite{andrews_est}.

Assumption \ref{loss function} imposes convexity on the penalty
function. We do not consider non-convex penalty functions for
pragmatic reasons. One of the main motivations of this paper is
the generic computability of the estimates, given that they solve
well-defined convex optimization problems.  The domination
condition, $\rho(h) \leq  1 + | h |^p$ for some $1 \leq p <
\infty$, is conventional and is satisfied in all examples of
$\rho$ we gave.

The assumption that $\varphi(\xi)= \int \rho(u-\xi) e^{\sss -u'a
u} du \propto E \rho( \mathcal{N}(0, a^{-1}) -\xi) $ attains a
unique minimum at some finite $\xi^*$ for any positive definite
$a$ is required, and it clearly holds for all of examples of
$\rho$ we mentioned. In fact, when $\rho$ is symmetric, $\xi^*=0$
by \cite{anderson_lemma}'s lemma.

Assumption \ref{consistency assumption} is implied by the usual
uniform convergence and  unique identification conditions as in
\cite{amen}. The proof of Lemma \ref{amen consistency}  can be
found in \cite{amen} and \cite{white:QML}.

\begin{lemma}\label{amen consistency} Given Assumption 1,
Assumption \ref{consistency assumption} holds if  there is a
 function $M_n(\theta)$ that
\begin{itemize}
\item[\textbf{i.}]
is nonstochastic, continuous on $\Theta$, for any $\delta > 0$,
$\limsup_n ( \sup_{|\theta - \theta_0|>\delta} M_n(\theta) -
M_n(\theta_0) ) < 0$,
 \item[\textbf{ii.}] $ L_n\(\theta\)/n -M_n(\theta) $
  converges to zero in (outer) probability uniformly over
$\Theta$.
\end{itemize}
\end{lemma}

Assumption 4 is satisfied under the conditions of Lemma 2, which
are known to be mild in nonlinear models. Assumption \ref{second
order expansion}.ii requires asymptotic normality to hold, and is
generally a weak assumption for cross-sectional and many
time-series applications. Assumption \ref{second order
expansion}.iii rules out the cases of mixed asymptotic normality
for some non-stationary time series models (which can be
incorporated at a notational cost with different scaling rates).

 Assumption
\ref{second order expansion}.iv easily holds when there is enough
smoothness.

\begin{lemma}\label{amemiya} Given Assumptions 1 and 3,
Assumption \ref{second order expansion} holds with
$$
\Delta_n(\theta_0) = \nabla_{\theta} L_n(\theta_0) \text{ and }
J_n(\theta_0) = - \nabla_{\theta \theta'} M_n\(\theta_0\) =O(1),
\text{ if }
$$
\begin{itemize}\item[\textbf{i.}] for some $\delta
> 0$, $L_n\(\theta\)$ and $M_n\(\theta\)$ are twice continuously differentiable in $\theta$ when
$\vert \theta - \theta_0\vert < \delta$,  \item [\textbf{ii.}]
there is $\Omega_n(\theta_0)$ such that $\Omega_n^{\sss
-1/2}(\theta_0)\nabla_{\theta} L_n(\theta_0)/\sqrt{n} \ \ind
\mathcal{N}(0, I)$, \ \ $J_n(\theta_0)=O(1)$ and
$\Omega_n(\theta_0) =O(1)$ are uniformly positive definite, and
\item[\textbf{iii.}]  for some $\delta>0$ and each $\epsilon >0$
$$\limsup_{n \rightarrow \infty} P^*\left \{ \sup_{|
\theta -\theta_0| < \delta}  | \nabla_{\theta \theta'}
L_n\(\theta\)/n -
 \nabla_{\theta \theta'} M_n\(\theta\)
 | > \epsilon\right \} =0.$$   \end{itemize}
\end{lemma}
Lemma 2 is immediate, hence its proof is omitted. Both Lemmas 1
and 2 are simple but useful conditions that can be easily verified
using standard uniform laws of large numbers and central limit
theorems. In particular, they have been proven to hold for
criterion functions corresponding to
\begin{itemize}
\item[\textbf{1.}] Most smooth cross-sectional models described in
\cite{amen};
%\item[\textbf{2.}] The smooth GMM framework
%for stationary data satisfying the Gordin (mixingale type)
%conditions, of \cite{hansen82};
\item[\textbf{2.}] The smooth nonlinear stationary and dynamic GMM and Quasi-likelihood
models of \cite{hansen:gmm}, \cite{gallant:white} and
\cite{potscher}, covering Gordin(mixingale type) conditions and
near-epoch dependent processes such as ARMA, GARCH, ARCH, and
other models alike;
\item[\textbf{3.}] General empirical likelihood models for smooth moment equation models
 studied by \cite{imbens:el}, \cite{kitamura_stutzer}, \cite{neweysmith},
Owen
(\citeyear{owen:1989},\citeyear{owen:1990},\citeyear{owen:1991},
\citeyear{owen_book}), \cite{qin_lawless}, and the recent
extensions to the conditional moment equations.

\end{itemize}

Hence the  main results of this paper, Theorems 1-4, apply to
these fundamental econometric and statistical models. Moreover,
Assumption \ref{second order expansion} does not require
differentiability of the criterion function and thus holds even
more generally. Assumption \ref{second order
expansion}.\textbf{iv} is a Huber-like stochastic equicontinuity
condition, which requires that the remainder term of the
expansion can be controlled in a particular way over a
neighborhood of $\theta_0$. In addition to Lemma 2, many
sufficient conditions  for Assumption 4 are given in empirical
process literature, e.g. \cite{amen}, \cite{andrews1},
\cite{newey:eq}, \cite{pakes_pollard}, and \cite{vaart}. Section
4 verifies Assumption 4 for the leading models with nonsmooth
criterion functions, including the examples discussed in the
previous section.

\subsection{Convergence in the Total Variation of Moments Norm}
Under Assumptions 1-4, we show that the Quasi-posterior density
concentrates around $\theta_0$ at the speed  $1/\sqrt{n}$ as
measured by the total variation of moments norm, and then use
this preliminary result to prove all other main results.

Define the local parameter $h$ as a normalized deviation from
$\theta_0$ and centered at the normalized random ``score
function": \bs h = \sqrt{n}\(\theta - \theta_0\) -
J_n\(\theta_0\)^{-1} \Delta_n\(\theta_0\)/\sqrt{n}.
\end{split}\end{align}
Define by the Jacobi rule the localized Quasi-posterior density
for $h$ as \bs p_n^*\(h\) = \frac{1}{\sqrt{n}}
p_n\(h/\sqrt{n}+\theta_0+J_n\(\theta_0\)^{-1}
\Delta_n\(\theta_0\)/n\).
\end{split}\end{align}
Define the total variation of moments norm for a real-valued
measurable function $f$ on $S$ as
$$
\| f \|_{\sss TVM(\alpha)} \equiv \int_{S} ( 1 + |h|^{\alpha})
|f(h)| dh.
$$

\vspace{.1in}
\begin{theorem}[\textbf{\textit{Convergence in Total  Variation of Moments Norm}}]\label{theorem2} Under Assumptions
\ref{parameter} - \ref{second order expansion}, for any $0 \leq
\alpha < \infty$, \bs
 & \| p_n^*(h) -   p^*_{\infty}(h)  \|_{\sss TVM(\alpha)} \equiv \int_{H_n}  (1 + | h |^\alpha)  |
p_n^*(h) -   p^*_{\infty}(h)   | dh \inp 0,
\end{split}\end{align}
where $H_n = \{ \sqrt{n}(\theta - \theta_0)- J_n\(\theta_0\)^{-1}
\Delta_n\(\theta_0\)/\sqrt{n}: \theta \in \Theta \}$ and $$
 p^*_{\sss \infty}(h) = {\sqrt{\scriptstyle \frac{\det J_n(\theta_0)}{(2
 \pi)^d}}} \cdot \exp \( - \frac{1}{2} h'J_n(\theta_0)h\).
$$\end{theorem}
\vspace{.1in}

Theorem \ref{theorem2} shows that $p_n\(\theta\)$ is concentrated
at a $1/\sqrt{n}$ neighborhood of $\theta_0$ as measured by the
total variation of moments norm. For large $n$, $p_n\(\theta\)$ is
approximately a random normal density with the random mean
parameter $\theta_0+ J_n\(\theta_0\)^{-1} \Delta_n\(\theta_0\)/n,$
and constant variance parameter $J_n(\theta_0)^{-1}/n.$

Theorem \ref{theorem2} applies to general statistical criterion
functions $L_n(\theta)$, hence it covers the parametric likelihood
setting as a fundamental case, in particular implying the
Bernstein-Von Mises theorems, which state the convergence of the
likelihood posterior to the limit random density in the total
variation norm.  Note also that the total variation norm results
from setting $\alpha =0$ in the total variation of moments norm.
The use of the latter is needed to deduce the convergence of
LTE's such as the posterior means or variances in Theorems 2-4.

\subsection{Limit Results for Point Estimates and Confidence Intervals}
As a consequence of Theorem \ref{theorem2}, Theorem 2 establishes
$\sqrt{n}$- consistency and asymptotic normality of LTE's. When
the loss function $\rho\(\cdot\)$ is symmetric, LTE's are
asymptotically equivalent to the extremum estimators.

Recall that the extremum estimator $\sqrt{n} ( \widehat
\theta_{ex} - \theta_0)$, where $\widehat \theta_{ex} = \arg
\sup_{\theta \in \Theta} L_n(\theta)$, is first order-equivalent
to
$$
U_n \equiv \frac{1}{\sqrt{n}}
J_n(\theta_0)^{-1}\Delta_n(\theta_0).
$$
Given that the $p_n^*$ approaches $p^*_{\sss \infty}$, it may be
expected that the LTE $\sqrt{n}(\widehat \theta - \theta_0)$ is
asymptotically equivalent to
$$
Z_n = \arg\inf_{z \in \Bbb{R}^d} \left \{ \ \ \int_{\Bbb{R}^d}
\rho\(z-u\) p^*_{\sss \infty} ( u - U_n  ) \ du \ \right \}.
$$
To see a relationship between $Z_n$ and $U_n$, define
$$
\xi_{\sss J_n(\theta_0)} \equiv \arg\inf_{z \in \Bbb{R}^d} \left
\{ \ \ \int_{\Bbb{R}^d} \rho\(z-u\) p^*_{\sss \infty} (u) \ du \
\right \},
$$
which exists by Assumption 2.\footnote{For example, in the scalar
parameter case, if $\rho(h) = (\alpha - 1(h <0))h$, the constant
$\xi_{\sss J(\theta_0)} = q_{\alpha} J_n(\theta_0)^{-1/2}$, where
$q_{\alpha}$ is the $\alpha$-quantile of $\mathcal{N}(0,1)$.} If
$\rho$ is symmetric, i.e. $\rho(h) = \rho(-h)$, then by
Anderson's lemma $ \xi_{\sss J_n(\theta_0)} = 0. $ Hence
$$
Z_n = \xi_{\sss J_n(\theta_0)} + U_n,
$$
and we are prepared to state the result.

\vspace{.1in}
\begin{theorem}[\textbf{\textit{LTE in Large Samples}}]\label{theorem3}
 Under Assumptions \ref{parameter}-\ref{second order expansion},
  \bs
 & \sqrt{n}(\hat\theta - \theta_0)  =
\xi_{\sss J_n(\theta_0)} + U_n +o_p(1), \ \
 \Omega_n^{\sss -1/2}(\theta_0) J_n(\theta_0) U_n \ind
\mathcal{N}(0 \ , I ).
 \end{split}\end{align}
Hence $$\Omega_n^{\sss -1/2}(\theta_0) J_n(\theta_0) \(
\sqrt{n}(\hat\theta - \theta_0) - \xi_{\sss J_n(\theta_0)}\) \ind
\mathcal{N}( 0, I).
$$ If loss function $\rho_n$ is symmetric, i.e. $\rho_n\( h\) = \rho_n(-h)$ for all
$h$, $\xi_{\sss J_n(\theta_0)}=0$ for each $n$.
\end{theorem}
\vspace{.1in}

In order for the Quasi-posterior distribution to provide valid
large sample confidence intervals, the density of
$W_n=\mathcal{N}\(0, J_n(\theta_0)^{-1} \Omega_n(\theta_0)
J_n(\theta_0)^{-1}\)$ should coincide with that of $p^*_{\sss
\infty}\(h\)$. This requires \bs & \int_{\Bbb{R}^d} h h' p^*_{\sss
\infty} (h) d h \equiv J_n(\theta_0)^{-1} \sim \text{Var} ( W_n)
\equiv J_n(\theta_0)^{-1} \Omega_n(\theta_0) J_n(\theta_0)^{-1},
\end{split}\end{align}
or equivalently
$$
\Omega_n(\theta_0) \sim J_n(\theta_0),
$$
which is a  \textbf{generalized information equality.} The
information equality is known to hold for regular, correctly
specified likelihoods. It is also known to hold for appropriately
constructed criterion functions of generalized method of moments,
minimum distance estimators, generalized empirical likelihood
estimators, and properly weighted extremum estimators; see
Section 4.

Consider construction of the confidence intervals for the quantity
$ g(\theta_0)$, and suppose $g$ is continuously differentiable.
Define \bs F_{g,n}( x) = \int_{\theta \in \Theta: g(\theta) \leq
x } p_n (\theta) d \theta,\quad\text{and}\quad c_{g,n}(\alpha) =
\inf \{ x : F_{g,n}( x) \geq \alpha \}.
\end{split}\end{align}
Then a LT confidence interval is given by $\[ c_{g,n}(\alpha/2) ,
c_{g,n}(1-\alpha/2)   \].$ As previously mentioned,  these
confidence intervals can be constructed by using the $\alpha/2$
and $1-\alpha/2$ quantiles of the MCMC sequence
$$
\( g(\theta^{(1)}), ..., g(\theta^{\sss (B)})\)
$$
 and thus
are quite simple in practice.  In order for the intervals to be
valid in large samples,  one needs to ensure the generalized
information equality,  which can be done easily through the use of
optimal weighting  in GMM and minimum-distance criterion functions
or  the use of generalized empirical likelihood functions; see
Section 4.

 Consider now the
usual asymptotic intervals based on the $\Delta$- method and any
estimator with the property \bs \sqrt n (\widehat \theta -
\theta) = J_n(\theta_0)^{-1} \Delta_n(\theta) / \sqrt{n}+ o_p(1).
\end{split}\end{align}
Such intervals are usually given by \bs
\[ g(\widehat \theta) + q_{\alpha/2}{\sss \frac{\sqrt{ \nabla_{\theta}
g(\widehat \theta)^{'} J_n(\theta_0)^{-1}\nabla_{\theta}g(\widehat
\theta) }}{\sqrt{n}}},  \ \ \ g(\widehat \theta) +
q_{1-\alpha/2}{\sss \frac{\sqrt{ \nabla_{\theta} g(\widehat
\theta)^{'}\widehat J_n( \theta_0)^{-1}\nabla_{\theta} g(\widehat
\theta) }}{\sqrt{n}}} \],
\end{split}\end{align}
where $q_{\alpha}$ is the $\alpha$-quantile of the standard normal
distribution. The following theorem establishes the large sample
correspondence of the Quasi-posterior confidence intervals to the
above intervals.

\begin{theorem}[\textit{\textbf{Large Sample Inference I}}]\label{information matrix}
Suppose Assumptions 1-4 hold. In addition suppose that the
generalized information equality holds: \bs \lim_{n \to \infty}
J_n( \theta_0) \Omega_n(\theta_0)^{-1} = I.
\end{split}\end{align}
Then for any $\alpha \in (0,1)$  $$c_{g, n}(\alpha)  - g(\widehat
\theta) - q_{\alpha}{\displaystyle \frac{\sqrt{ \nabla_{\theta} g(
\theta_0)^{'} J_n(\theta_0)^{-1}\nabla_{\theta}g(\theta_0)
}}{\sqrt{n}}}= o_p\(\frac{1}{\sqrt{n}}\),$$ and \bs \lim_{n
\rightarrow \infty} P^* \Big \{ c_{g, n}(\alpha/2) \leq
g(\theta_0) \leq
c_{g, n}(1-\alpha/2) \Big \} = 1-\alpha. \\
\end{split}\end{align}
\end{theorem}

One practical limitation of this result arises in the case of
regression criterion functions (M-estimators), where achieving the
information equality may require nonparametric estimation of
appropriate weights,  e.g. as in censored quantile regression
discussed in Section 4. This may entirely be avoided by using a
different method for construction of confidence intervals.
Instead of the Quasi-posterior quantiles, we can use the
Quasi-posterior variance as an estimate of the inverse of the
population Hessian matrix $J^{-1}_n(\theta_0)$, and combine it
with any available estimate of $\Omega_n(\theta_0)$ (which
typically is easier to obtain) in order to obtain the
$\Delta$-method style intervals. The usefulness of this methods
is particularly evident in the censored quantile regression,
where direct estimation of $J_n(\theta_0)$ requires use of
nonparametric methods.

\begin{theorem}[\textit{\textbf{Large Sample Inference II}}]\label{information matrix fails}
Suppose Assumptions 1-4 hold. Define for $\widehat \theta =
\int_{\Theta} \theta p_n(\theta) d \theta$,
$$
\widehat J_n^{-1}( \theta_0) \equiv  \int_{\Theta} n (\theta-
\widehat \theta )(\theta - \widehat \theta)' p_n(\theta) d \theta,
$$
and
$$c_{g,n}(\alpha)\equiv g(\widehat \theta)
+ q_{\alpha}\cdot {\displaystyle \frac{\sqrt{ \nabla_{\theta} g(\widehat
\theta)^{'}\widehat J_n(\theta_0)^{-1} \widehat \Omega_n
(\theta_0) \widehat J_n(\theta_0)^{-1} \nabla_{\theta}g(\widehat
\theta) }}{\sqrt n}},$$ where $\widehat \Omega_n(\theta_0)
\Omega_n^{-1}(\theta_0) \inp I$. Then $\widehat
J_n(\theta_0)J_n(\theta_0)^{-1} \inp I$, and \bs \lim_{n
\rightarrow \infty} P_* \Big \{ c_{g, n}(\alpha/2) \leq
g(\theta_0) \leq c_{g, n}(1-\alpha/2) \Big \} = 1-\alpha.
\end{split}\end{align}
\end{theorem}
 In practice $\widehat J_n( \theta_0)^{-1}$ is computed by
multiplying by $n$ the variance-covariance matrix of the MCMC
sequence  $ S = \(\theta^{(1)}, \theta^{(2)}, ..., \theta^{({\sss
B})} \) $.

\vspace{.1in}

\textbf{Remark} (Sandwich; added in 2022): Note that the above result
establishes that should use the Huber sandwich $$ \widehat J_n(\theta_0)^{-1} \widehat \Omega_n
(\theta_0) \widehat J_n(\theta_0)^{-1}$$ to perform frequentist inference using Quasi-Bayesian methods (and Bayesian methods) when the generalized information equality does not hold.  Several more recent papers arrive at a similar point. $\square$

\section{Applications to Selected Problems}

This section further elaborates the approach through several
examples. Assumptions 1-4 cover a wide variety of smooth
econometric models (by virtue of Lemma 1 and Lemma 2). Thus, what
follows next is mainly motivated by models with non-smooth moment
equations, such as those occurring in Examples 1 - 3.
Verification of the key Assumption 4 is not immediate in these
examples, and Propositions 1-3 and the forthcoming examples show
how to do this in a class of models that are of prime interest to
us.

\subsection{Generalized Method of Moments and Nonlinear Instrumental Variables}\label{gmm section}
Going back to Example 2, recall that a typical model that
underlies the applications of GMM is a set of population moment
equations: \bsnumber\label{gmm1} E m_i(\theta)
=0\quad\quad\quad\text{if and only if}\quad\quad \theta =
\theta_0.
 \end{split}\end{align}
Method of moment estimators involve maximizing an objective
function of the form \begin{eqnarray} \label{gmm2}
  L_n\(\theta\) & = &- n \( g_n\(\theta\) \)' W_n(\theta)
  \(g_n\(\theta\)\)/2,
  \\ g_n\(\theta\) &= & \frac{1}{n}\sum_{i=1}^n m_i\(\theta\), \\
  \label{gmm3}
  W_n(\theta)  & = & W(\theta) +o_p(1)\ \text{ uniformly in } \theta \in
  \Theta,
  \\
   \label{gmm4}
  W\ (\theta) & >& 0 \text{ and continuous  uniformly in } \theta \in
  \Theta, \\ \label{gmm5}
  W(\theta_0) & =& \[\lim_{n \rightarrow \infty} \text{Var} [ \sqrt{n}
  g_n(\theta_0)]\]^{-1}.
\end{eqnarray} The choice (\ref{gmm5}) of the weighting matrix implies
the generalized information equality under standard regularity
conditions.

Generally, by a Central Limit Theorem $\sqrt{n} g_n(\theta_0) \ind
\mathcal{N}(0, W^{-1}(\theta_0))$, so that the objective function
can be interpreted as the approximate log-likelihood for the
sample moments of the data $g_n(\theta)$. Thus we can think of
GMM as an approach that specifies an approximate likelihood for
selected moments of the data without specifying the likelihood of
the entire data.\footnote{This does not help much in terms of
providing formal asymptotic results for the GMM model.}

We may impose Assumptions 1-4 directly on the GMM objective
function.  However, to highlight the plausibility and elaborate on
some examples that satisfy Assumption 4 consider the following
proposition.

\begin{proposition}[\textbf{\textit{Method-of-Moments and Nonlinear
IV}}]\label{prop1} Suppose that Assumptions 1-2 hold, and that
for all $\theta$ in $\Theta$, $m_i(\theta)$ is stationary and
ergodic, and
\begin{itemize}
 \item[\textbf{i.}] conditions (\ref{gmm1})-(\ref{gmm4}) hold,
 \item[\textbf{ii.}] $J(\theta) \equiv G(\theta)' W(\theta) G(\theta) >0 $ and is continuous,
 $G(\theta) = \nabla_{\theta} E m_i(\theta)$ is continuous,
 \item[\textbf{iii.}] $\Delta_n(\theta_0)/\sqrt{n} =  -\sqrt{n} g_n(\theta_0)' W(\theta_0)
G(\theta_0) \ind \mathcal{N}(0,  \Omega(\theta_0))$,
$\Omega(\theta_0) \equiv G(\theta_0)' W(\theta_0) G(\theta_0)$,
\item[\textbf{iv.}]
for any $\epsilon > 0$, there is $\delta>0$ such that \bsnumber
\label{huber} \limsup_{n \rightarrow \infty} P^*\left \{
\sup_{|\theta - \theta'| \leq \delta} \frac{\sqrt{n} |
\(g_n(\theta) - g_n(\theta')\) - \(E g_n(\theta) - Eg_n(\theta')\)
| }{1 + \sqrt{n} | \theta - \theta' |} > \epsilon \right\} <
\epsilon.
 \end{split}\end{align}
\end{itemize}
Then Assumption \ref{second order expansion} holds. In addition
the information equality holds by construction. Therefore the
conclusions of Theorems \ref{theorem2}-\ref{information matrix
fails} hold with $\Delta_n(\theta_0)$, $\Omega_n(\theta_0)\equiv
\Omega\(\theta_0\)$ and $J_n(\theta_0)\equiv J\(\theta_0\)$
defined above, where the condition (\ref{gmm5}) is only needed for
the conclusions of Theorem \ref{information matrix} to hold.
\end{proposition}

Therefore, for symmetric loss functions $\rho_n$, the LTE is
asymptotically equivalent to the GMM extremum estimator.
Furthermore, the generalized information equality holds by
construction, hence Quasi-posterior quantiles provide a
computationally attractive method of ``inverting" the objective
function for the confidence intervals.

For twice continuously differentiable smooth moment conditions,
the smoothness conditions on $\nabla_{\theta} L_n(\theta) $ and
$\nabla_{\theta \theta'} L_n (\theta)$ stated in Lemma
\ref{amemiya} trivially imply condition \textbf{iv} in
Proposition \ref{prop1}. More generally, \cite{andrews1},
\cite{pakes_pollard} and \cite{vaart} provide many methods to
verify that condition  in a wide variety of method-of-moments
models.

\paragraph{\bf Example 3 Continued.} Instrumental median regression falls outside of both the
classical Bayesian approach and the classical smooth nonlinear IV
approach of \cite{amemiya:IV}. Yet the conditions of Proposition
\ref{prop1} are satisfied under mild conditions:
\begin{itemize}
\item[\textbf{i.}] $(Y_i, D_i, X_i, Z_i)$ is an iid data sequence,   $E
[m_i (\theta_0)Z_i] = 0$, and $\theta_0$ is identifiable,
\item[\textbf{ii.}] $\{ m_i(\theta) = \( \tau - 1( Y_i \leq q(D_i, X_i,
\theta))\) Z_i, \theta \in \Theta \}$ is a Donsker
class,\footnote{This is a very weak restriction on the function
class, and is known to hold for all practically relevant
functional forms, see \cite{vandervaart}. } $E \sup_{\theta} |
m_i(\theta) |^2 < \infty$,
\item[\textbf{iii.}] $ G(\theta) =
\nabla_{\theta} E m_i(\theta)= - E f_{\sss Y|D,X,Z}(q(D,X,
\theta)) Z \nabla_{\theta} q(D,X, \theta)'$
 is
continuous,
\item[\textbf{iv.}]  $ J(\theta) = G(\theta)'W(\theta)G(\theta)$ $>0$ and is continuous
in an open ball at $\theta_0$.
\end{itemize}
In this case the weighting matrix can be taken as \bs
 W_n(\theta) =  \frac{1}{\tau (1- \tau)}\[\frac{1}{n} \sum_{i=1}^n
Z_i Z_i'\]^{-1},
\end{split}\end{align}
so that the information equality holds. Indeed, in this case
$$
\Omega(\theta_0) = G(\theta_0)'W(\theta_0) G(\theta_0) =
J(\theta_0),
$$
where $$ W(\theta_0) = \text{plim} \ W_n(\theta_0)= \[\text{Var} \
m_i(\theta_0) \]^{-1}, \ \ \text{Var} \ m_i(\theta_0) = \tau (1-
\tau) E Z_i Z_i'.$$

When the model $q$ is linear and the dimension of $D$ is small,
there are computable and practical estimators in the
literature.\footnote{These include e.g. the ``inverse" quantile
regression approach in \cite{iqr}, which is an extension of
\cite{koenker:1978}'s quantile regression to endogenous
settings.} In more general models, the extremum estimates are
quite difficult to compute, and the inference faces the well-known
difficulty of estimating sparsity parameters.

On the other hand, the Quasi-posterior median and quantiles are
easy to compute and provide asymptotically valid confidence
intervals. Note that the inference does not require the
estimation of the density function.  The simulation example given
in Section 5 strongly supports this alternative approach.

Another important example which poses computational challenge is
the estimation problem of \cite{blp1}.  This example is similar
in nature to the instrumental quantile regression, and the
application of the LT methods may be fruitful there.

\subsection{Generalized Empirical Likelihood} A
class of objective functions that  are first-order equivalent to
optimally weighted GMM (after recentering) can be formulated using
the generalized empirical likelihood framework.

 A class of
generalized empirical likelihood  functions(GEL) are studied in
\cite{imbens_spady_johnson}, \cite{kitamura_stutzer}, and
\cite{neweysmith}. For a set of moment equations {$ E
m_i(\theta_0) = 0$} that satisfy the conditions of section
\ref{gmm section}, define \bsnumber\label{gel1} \bar L_n( \theta,
\gamma)
 \equiv
 \sum_{i=1}^n \( s\(m_i(\theta)' \gamma\) - s(0) \).
 \end{split}\end{align}
Then set \bsnumber\label{GEL1} L_n\(\theta\) = \bar L_n\(\theta,
\hat\gamma\(\theta\)\),\end{split}\end{align} where $ \widehat
\gamma (\theta)$ solves \bsnumber\label{inner saddle point}
\widehat \gamma(\theta) \equiv \arg \inf_{\gamma \in \Bbb{R}^p}
\bar L_n (\theta, \gamma),
\end{split}\end{align}
and $p = \dim( m_i)$.

\noindent The scalar function $s(\cdot)$ is a strictly convex,
finite, and three times differentiable function on an open
interval of $\Bbb{R}$ containing $0$, denoted $\mathcal V$,  and
is equal to $+ \infty$ outside such an interval. $s\(\cdot\)$ is
normalized so that both $\nabla s\(0\)=1$ and $\nabla^2 s\(0\) =
1$. The choices of the function $s(v) = - \ln (1-v)$, $\exp(v)$,
and  $(1+v)^2/2$ lead to the well-known empirical likelihood,
exponential tilting, and continuous-updating GMM criterion
functions.

Simple and practical sufficient conditions for  Lemma
\ref{amemiya} are given in \cite{qin_lawless},
\cite{imbens_spady_johnson}, \cite{kitamura:annals},
\cite{kitamura_stutzer}, including stationary weakly dependent
data, \cite{neweysmith},  and \cite{hahnexp}. Thus, the
application of LTE's to these problems is immediate.

To illustrate  a further use of LTE's we state a set of simple
conditions geared towards non-smooth microeconometric applications
such as the instrumental quantile regression problem. These
regularity conditions imply the first order equivalence of the
GEL and GMM objective functions. The Donskerness condition below
is a weak assumption that is known to hold for all reasonable
linear and nonlinear functional forms encountered in practice, as
discussed in \cite{vandervaart}.

\begin{proposition}[\textbf{\textit{Empirical Likelihood Problems}}] Suppose that Assumptions
1- 2 hold, and that the following conditions are satisfied: for
some $\delta>0$ and all $\theta \in \Theta$
\begin{itemize}
\item[\textbf{i.}] condition (\ref{gmm1}) holds and that $m_i(\theta)$ is iid,
%\item[\textbf{ii.}] $\partial P[m_i(\theta)< x]/ \partial \theta$ is
% continuous in $\theta$ uniformly in $x:|x| \leq K$, for $K$ in
% \textbf{iii.}
\item[\textbf{ii.}] $E m_i(\theta)$ is continuously differentiable in $\theta$ with derivative $G(\theta)$, and $V(\theta)=E m_i(\theta) m_i(\theta)'$ is continuous in $\theta$; with $G(\theta_0)$ having full rank;
\item[\textbf{iii.}] $\sup_{|\theta - \theta_0|<\delta} |m_i (\theta)| < K$
a.s., for some constant $K$

\item[\textbf{iv.}] $\{ m_i (\theta), \theta \in \Theta\}$ is
Donsker class, where $$\sqrt{n} g_n(\theta_0) =
\frac{1}{\sqrt{n}}\sum_{i=1}^n m_i\(\theta_0\) \ind
\mathcal{N}\(0, V\(\theta_0\)\), \ \ V\(\theta_0\) = E \{
m_i(\theta_0)   m_i(\theta_0)'\}>0,$$
\end{itemize}
then Assumptions \ref{consistency assumption} and \ref{second
order expansion} hold, and thus the conclusions of Theorems
\ref{theorem2} - \ref{information matrix fails} are true with \bs
&\Delta_n\(\theta_0\) / \sqrt{n} = \sqrt{n} g_n(\theta_0)
V(\theta_0)^{-1} G(\theta_0) \ind \mathcal{N}( 0,
\Omega(\theta_0)), \\
 &\Omega(\theta_0) = G(\theta_0)'  V(\theta_0)^{-1}  G(\theta_0),   \\
&J(\theta_0) = G(\theta_0)'  V(\theta_0)^{-1}  G(\theta_0),
 \ \ G(\theta_0) =  \nabla_{\theta}  E m_i(\theta_0).
\end{split}\end{align}
The information equality holds in this case.
\end{proposition}

\textbf{Remark.} We have corrected condition ii. relative to the published version. In particular, the full rank condition $G(\theta_0)$ was implicitly used in the proof but missing in the statement of assumptions.  $\square$

Another (equivalent) way to proceed is through the dual
formulation. Consider the following criterion function
 \bsnumber\label{Dual}
L_n(\theta) = \sup_{\pi_1,..., \pi_n \in [0,1] } \sum_{i=1}^n h (
\pi_i) \text{ subject to }  \sum_{i=1}^n m_i(\theta) \pi_i = 0,
\sum_{i=1}^n \pi_i =1,
 \end{split}\end{align}
where $h$ is the Cressie-Reid divergence criterion, cf.
\cite{neweysmith}
$$
h(\pi) = \frac{1}{\gamma(\gamma +1)}
\frac{(\frac{\pi}{1/n})^{\gamma+1} -1}{n}.
$$
The function $L_n(\theta)$ in (\ref{Dual}) is the generalized
empirical likelihood function for $\theta$ with the concentrated
out probabilities.  In fact, (\ref{Dual}) corresponds to
(\ref{GEL1}) by the argument given in \cite{qin_lawless}
p.303-304, so that Proposition 2 covers (\ref{Dual}) as a special
case up to renormalization.

Empirical probabilities $\widehat \pi_i(\theta)$'s are obtained in
(\ref{Dual}) using the extremum method. The case $\gamma = -1$
yields the empirical likelihood case, where $\widehat
\pi_i(\theta)$'s are obtained through the maximum likelihood
method. Taking $\gamma=0$ yields the exponential tilting case,
where $\widehat \pi_i(\theta)$'s are obtained through
minimization of the Kullback-Leibler distance from the empirical
distribution. Taking $\gamma=1$ yields the continuous-updating
case, where $\widehat \pi_i(\theta)$'s are obtained through the
minimization of the Euclidean distance from the empirical
distribution. Each approach generates the implied probabilities
$\widehat \pi_i(\theta)$ given $\theta$. \cite{qin_lawless} and
\cite{neweysmith} provide the formulas: $$ \widehat \pi_i(\theta)
= \nabla s(\widehat \gamma(\theta)' m_i (\theta))/\sum_{i=1}^n
\nabla s(\widehat \gamma(\theta)' m_i (\theta)).$$

The Quasi-posterior for $\theta$ and $\widehat \pi_i(\theta)$ can
be used for predictive inference. Suppose $m_i(\theta) = m(X_i,
\theta)$ for some random vector $X_i$. Then the Quasi-posterior
predictive probability is given by
 \bsnumber\label{predictive1}
\widehat P\{ X_i \in A\} = \int  \underbrace{\sum_{i=1}^n \widehat
\pi_i (\theta) 1\{X_i \in A\}}_{\equiv h_n(\theta)} p_n(\theta)
d\theta = \int h_n(\theta) p_n(\theta) d \theta,
 \end{split}\end{align}
which can be computed by averaging over the MCMC sequence
evaluated at $h_n$, $ \(h_n(\theta^{(1)}),...,
h_n(\theta^{(B)})\).$ It follows similarly to the proof of Theorem
1 in \cite{qin_lawless} that
 \bsnumber\label{predictive2}
\sqrt{n} ( \widehat P\{ X_i \in A\} - P\{X_i \in A\}) \ind
\mathcal{N}\(0, \Pi_{\sss A}\),
  \end{split}\end{align}
where  $\Pi_{\sss A} \equiv  P\{X_i \in A\} (1 - P\{X_i \in A\}) -
E m_i(\theta_0)' 1\{X_i \in A\} \cdot U \cdot E m_i(\theta_0)
1\{X_i \in A\} $,  $U = V(\theta_0)^{-1} \{ I - G(\theta_0)
J(\theta_0)^{-1} G(\theta_0)V(\theta_0)^{-1} \}$.

\subsection{M-estimation}
M-estimators, which include many linear and nonlinear regressions
as special cases, typically maximize objective functions of the
form
$$L_n\(\theta\) = \sum_{i=1}^n m_i\(\theta\).$$
$m_i\(\theta\)$ need not be the log likelihood function of
observation $i$, and may depend on preliminary non-parametric
estimation. Assumptions \ref{parameter}--\ref{consistency
assumption} usually are satisfied by uniform laws of large numbers
and by unique identification of the parameter; see for example
\cite{amen} and \cite{newey_mcfadden}. The next proposition gives
a simple set of sufficient conditions for Assumption \ref{second
order expansion}.

%\begin{proposition}[\textbf{\textit{M- problems}}] Suppose
%assumptions 1-3 hold for the criterion function specified above
%with the following additional properties: over an open ball at
%$\theta_0$
% \begin{itemize}
%\item[(i)] $m_i(\theta)$ is stationary, strongly  mixing sequence, \\
% \item[(ii)] $|m_i(\theta_1)- m_i(\theta_2)|\leq c_i | \theta_1 -
% \theta_2|$, $Ec_i^2 < \infty$,    $ J(\tau) = \nabla_{\theta \theta'} E[m_i(\theta)] $ is continuous and p.d.\\
% \item[(iii)] $\psi_i(\theta) = \nabla_{\theta} m_i(\theta),$ exists
% a.s.,  $n^{-1/2} \sum_{i=1}^n \psi_i(\theta)  \ind N(0, \Omega(\theta_0))$ \\
% \item[(iv)] $E [ m_i(\theta) - m_i (\theta_0) - \psi_i(\theta_0)'(\theta - \theta_0)]^2 = o( | \theta- \theta_0|^2)
% $
% \end{itemize}
%The conclusions of Theorems 1 and 2 hold, and if $J(\theta_0)^{-1}
%= \Omega(\theta_0)$ the conclusions of Theorem 3 holds.
%\end{proposition}
\begin{proposition}[\textbf{\textit{M-problems}}] \label{m-proposition}
Suppose Assumptions \ref{parameter}-\ref{consistency assumption}
hold for the criterion function specified above with the following
additional conditions: Uniformly in $\theta$ in an open
neighborhood of $\theta_0$, $m_i(\theta)$ is stationary and
ergodic, and for $\overline{m}_n(\theta)= \sum_{i=1}^n
m_i(\theta)/n$,
 \begin{itemize}
 \item[\textbf{i.}] there exists $\dot m_i(\theta_0)$ such that
 $E\dot m_i(\theta_0) =0$ for each $i$ and, for some $\delta >0$,
\bs &\bigg \{ \frac{m_i(\theta)- m_i(\theta_0) - \dot
m_i(\theta_0)' (\theta -
 \theta_0)}{| \theta - \theta_0 |},  \ \ \theta: | \theta - \theta_0 | < \delta \bigg \}
  \text{ is a Donsker class, }\\
& E [ \overline{m}_n(\theta) -\overline{m}_n (\theta_0) -
\overline{\dot m}_n(\theta_0)'(\theta - \theta_0)]^2 = o( |
\theta- \theta_0|^2), \\ & \Delta_n\(\theta_0\)/\sqrt{n} =
\sum_{i=1}^n \dot m_i(\theta_0)/\sqrt{n}  \ind \mathcal{N}(0,
\Omega(\theta_0)).
\end{split}\end{align}
\vspace{-.1in}
 \item[\textbf{ii.}]   $ J(\theta) = - \nabla_{\theta \theta'} E[m_i(\theta)] $ is continuous and nonsingular
 in a ball at
 $\theta_0$.
 \end{itemize}
Then Assumption 4 holds. Therefore, the conclusions of Theorems
\ref{theorem2}, \ref{theorem3}, and 4 hold. If in addition
$J(\theta_0) = \Omega(\theta_0)$, then the conclusions of Theorem
\ref{information matrix} also hold.
\end{proposition}

The above conditions apply to many well known examples such as
LAD, see for example \cite{vaart}. Therefore, for many nonlinear
regressions, Quasi-posterior means, modes, and medians are
asymptotically equivalent, and Quasi-posterior quantiles provide
asymptotically valid confidence statements if the generalized
information equality holds.  When the information equality fails
to hold, the method of Theorem 4 provides valid confidence
intervals.

\paragraph{\bf Example 1 Continued.}
Under the conditions given in \cite{powell_lad} or
\cite{newey:powell} for the censored quantile regression, the
assumptions of Proposition \ref{m-proposition} are satisfied.
Furthermore, it is not difficult to show that when the weights
$\omega^*_i$ are nonparametrically estimated, the conditions of
\cite{newey:powell} imply Assumption \ref{second order
expansion}. Under iid sampling, the use of efficient weighting
 \bsnumber\label{np_weights}
\omega^*_i = \frac{1}{\tau(1-\tau)}f_i,
 \end{split}\end{align}
where $f_i = f_{\sss Y_i|X_i}(q_i),  \ q_i = q(X_i;\theta_0),$
validates the generalized information equality, and the
Quasi-posterior quantiles form asymptotically valid confidence
intervals.  Indeed, since \bsnumber J(\theta_0) =
\frac{1}{\tau(1-\tau)}E  f_i^2 \nabla q_i \nabla q_i ', \text{
for } \nabla q_i =  \partial q(X_i, \theta_0)/\partial\theta,
\end{split}\end{align}
and
 \bsnumber  \frac{1}{\sqrt{n}}\Delta_n(\theta_0)= \frac{1}{\sqrt{n}}
\sum_{i=1}^n w_i \cdot (\tau - 1(Y_i \leq q_i)) \nabla q_i \ \ind
\ \mathcal{N}(0, \Omega(\theta_0)),\end{split}\end{align} with
 \bsnumber
\Omega(\theta_0) = \frac{1}{\tau(1-\tau)}E f_i^2 \nabla q_i
\nabla q_i ',
\end{split}\end{align}
we have
 \bsnumber\label{np_equality}
\Omega(\theta_0) = J(\theta_0).
 \end{split}\end{align}
For this class of problems, the Quasi-posterior means and medians
are asymptotically equivalent to the extremum estimators. The
Quasi-posterior quantiles provide asymptotically valid confidence
intervals when the efficient weights are used. However,
estimation of efficient weights requires preliminary estimation
of parameter $\theta_0$. When other weights are used, the method
of Theorem 4 provides valid confidence intervals.

\section{Computation and Simulation Examples}\label{simulation}
In this section we briefly discuss the MCMC method and present
simulation examples.

\subsection{Markov Chain
Monte Carlo }
%\label{computation_methods}

The Quasi-posterior density is proportional to {\bs p_n\(\theta\)
\propto
    e^{L_{\sss n}(\theta)} \pi(\theta).
\end{split}\end{align}}
In most cases we can easily compute $e^{L_{\sss n}(\theta)}
\pi(\theta)$. However, computation of the point estimates and
confidence intervals typically requires evaluation of  integrals
like \bsnumber\label{pos integral}
 \frac{\int_{\Theta} g(\theta) e^{L_{\sss n}(\theta)} \pi(\theta)
 d\theta} {\int_{\Theta} e^{L_{\sss n}(\theta)} \pi(\theta)
 d\theta}
\end{split}\end{align}
for various functions $g$. For problems for which no analytic
solution exists for (\ref{pos integral}), especially in high
dimensions, MCMC methods  provide powerful tools for evaluating
integrals like the one above. See for example
\cite{chib_handbook}, \cite{gewekehandbook}, and
\cite{montecarlo} for excellent treatments.

MCMC  is a collection of methods that produce an ergodic Markov
chain with the stationary distribution $p_n$. Given a starting
value $\theta^{(0)}$, a chain $(\theta^{(t)},  1 \leq t \leq B)$
is generated using a transition kernel with stationary
distribution $p_n$, which ensures the convergence of the marginal
distribution of $\theta^{(B)}$ to  $p_n$. For sufficiently large
$B$, the MCMC methods produce a dependent sample $(\theta^{(0)},
\theta^{(1)},..., \theta^{(B)})$ whose empirical distribution
approaches $p_n$. The ergodicity and construction of the chains
usually imply that as $B\rightarrow \infty$, \bs
 \frac{1}{B} \sum_{t=1}^B g (\theta^{(t)})  \inp \int_{\Theta} g(\theta)
 p_n(\theta) d \theta.
\end{split}\end{align}
We stress that this technique does not rely on the likelihood
principle and can be fruitfully used for computation of LTE's.
(Appendix B provides the formal details.)

One of the most important MCMC methods is the Metropolis-Hastings
algorithm.

\textbf{Metropolis-Hastings (MH) algorithm with Quasi-Posteriors}.
Given the Quasi-posterior density $p_n(\theta) \propto e^{L_{\sss
n}(\theta)} \pi(\theta)$, known up to a constant, and a
prespecified conditional density $q(\theta'|\theta)$, generate
$\(\theta^{(0)},...,\theta^{\sss (B)}\)$ in the following way,
 \begin{itemize}
 \item[1.] Choose a starting value $\theta^{(0)}$.
 \item[2.] Generate $\xi$ from $q(\xi|\theta^{(j)})$.
\item[3.]
 Update $\theta^{(j+1)}$ from $\theta^{(j)}$ for
 $j=1,2,...$, using
{\bs \theta^{(j+1)} =  \left \{ \begin{array}{ccc}
  \xi  &\text{ with probability }  &\rho(\theta^{(j)}, \xi)  \\
  \theta^{(j)}  &\text{ with probability } &1- \rho(\theta^{(j)}, \xi)
\end{array} \right .,
 \end{split}\end{align}}
where {\bs \rho(x,y) =  \min \( \frac{  e^{L_{\sss n}(y)} \pi(y)
q(x| y)} {  e^{L_{\sss n}(x)} \pi(x) q(y|x)} , 1 \right ).
\end{split}\end{align}}
 \end{itemize}

Note that the most important quantity in the algorithm is the
probability $\rho(x,y)$ of the move from an ``old" point $x$ to
the ``new" point $y$, which depends on how much of an improvement
in $e^{L_{\sss n}(y)} \pi(y)$ a possible ``new" value of $y$
yields relative to $e^{L_{\sss n}(x)} \pi(x)$ at the ``old" value
$x$. Thus, the generated chain of draws spends a relatively high
proportion of time in the higher density regions and a lower
proportion in the lower density regions. Because such proportions
of times are balanced in the right way, the generated sequence of
parameter draws has the requisite marginal distribution, which we
then use for computation of means, medians, and quantiles. (How
closely the sequence travels near the mode is not relevant.)

Another important choice is the transition kernel $q$, also
called the instrumental density. It turns out that a wide variety
of kernels yield Markov chains that converge to the distribution
of interest. One canonical implementation of the MH algorithm is
to take
$$
q (x|y) = f ( | x - y|),
$$
where $f$ is a density symmetric around $0$, such as the Gaussian
or the Cauchy density. This implies that the proposals $
\(\theta^{(j)}\)$ follow a random walk. This is the implementation we
used in this paper. \cite{chib_handbook}, \cite{gewekehandbook}
and \cite{montecarlo} can be consulted for important details
concerning the implementation and convergence monitoring of the
algorithm.

It is now worth repeating  that the main motivation behind the
LTE approach is based on its efficiency properties (stated in
sections 3 and 4) as well as computational attractiveness.
Indeed, the LTE approach is as efficient as the extremum
approach, but may avoid the computational curse of dimensionality
through the use of MCMC.  LTE's are
 typically means or quantiles
 of a Quasi-posterior distribution,
 hence can be computed (estimated) at the
 parametric rate $1/\sqrt{B}$,\footnote{Note that
 the rates are used
 for the informal motivation. We fix $d$ in the discussion, but
 the rate may typically increase linearly or polynomially in $d$ if
 $d$ is allowed to grow.  }
 where $B$ is the number of MCMC draws  (functional evaluations).
Indeed, under canonical
 implementations, the MCMC chains are geometrically mixing, so the
 rates of convergence are the same as under independent sampling.
 In contrast,  the
 extremum estimator (mode) is computed (estimated) by the MCMC and similar grid-based
 algorithms at the nonparametric rate
 $(1/B)^{\frac{p}{d+2p}}$, where $d$ is the parameter dimension and $p$ is the smoothness
 order of the objective function.

We used an optimistic tone regarding the performance of MCMC.
Indeed, in the problems we study, the objective functions have
numerous local optima, but all exhibit a well pronounced global
optimum. These problems are important, and therefore the good
performance of MCMC and the derived estimators are encouraging.
However, various pathological cases can be constructed, see
\cite{montecarlo}. Functions may have multiple separated global
modes (or approximate modes), in which case MCMC may require
extended time for convergence. Another potential problem is that
the initial draw $\theta^{(0)}$ may be very far in the tails of
the posterior $p_n(\theta)$.  In this case, MCMC may also take
extended time to converge to the stationary distribution. In the
problems we looked at, this may be avoided by choosing a starting
value based on economic considerations or other simple
considerations. For example, in the censored median regression
example, we may use the starting values based on an initial Tobit
regression. In the instrumental median regression, we may use the
two stage least squares estimates as the starting values.

\subsection{Monte Carlo Example 1: Censored Median Regression}

As discussed in Section 2, a large literature has been devoted to
the computation of Powell's censored median regression estimator.
In the simulation example reported below, we find that both in
small and large  samples with high degree censoring, the LT
estimation may be a useful alternative to the popular iterated
linear programming algorithm of \cite{buchinsky:phd}. The model we
consider is
 \bs
&Y^* =\beta_0 + X'\beta+ u,\\
&X \overset{d}= \mathcal{N}\(0,I_3\), \ \ u = X_1^2
\mathcal{N}\(0,1\), \ \ Y = \max\(0, Y^*\).
\end{split}\end{align}
The true parameter $(\beta_0, \beta_1, \beta_2, \beta_3   )$ is
$\(-6, 3, 3, 3\)$, which produces about $40\%$ censoring.

The LTE is based on the Powell's objective function $L_n\(\theta\)
= -\sum_{i=1}^n \vert Y_i - \max\(0, \beta_0 + X_i'\beta\) \vert.$
The initial draw of the MCMC series is taken to be the ordinary
least squares estimate, and other details are summarized in
Appendix B.

Table \ref{table1} reports the results. The number in parentheses
in the iterated linear programming (ILP) results indicates the
number of times that this algorithm converges to a local minimum
of $0$. The first row for the ILP reports the performance of the
algorithm among the subset of simulations for which the algorithm
does not converge to the local minimum at 0.  The second row
reports the results for all simulation runs, including those for
which the ILP algorithm does not move away from the local
minimum. The LTE's (Quasi-posterior mean and median) never
converge to the local minimum of $0$, and they compare favorably
to the ILP even when the local minima are excluded from the ILP
results, as can be seen from Table \ref{table1}. When the local
minima are included in the ILP results, LTE's do markedly better.

\begin{center}\textbf{[Table 1 goes here.]}\end{center}

\subsection{Monte Carlo Example 2: Instrumental Quantile Regression}

We consider a simulation example similar to that in
\cite{quantint}. The model is \bs
& Y = \alpha_0+D'\beta_0 + u, \ \ \  u =   \sigma(D) \epsilon, \\
& D = \exp \mathcal{N}(0, I_3), \  \ \epsilon = \mathcal{N}(0,1),
\ \ \sigma(D) = (1+ \sum_{i=1}^3 D_{(i)})/5.
\end{split}\end{align}
The true parameter $\(\alpha_0,\beta_0\)$ equals $0$, and we
consider the instrumental moment conditions \bs g_n(\theta) &=
\frac{1}{n} \sum_{i=1}^n ( \frac{1}{2} - 1 (Y_i \leq \alpha+
D'\beta)) Z, \ \ Z= (1, D),  \\ W_n(\theta)  &= \[\frac{1}{n}
\sum_{i=1}^n (\frac{1}{2} - 1 (Y_i \leq \alpha+ D'\beta))^2  Z_i
Z_i'\]^{-1}.
\end{split}\end{align}
In simulations, the initial draw of the MCMC series is taken to
be the ordinary least squares estimate, and other details are
summarized in Appendix B.

While instrumental median regression is designed specifically for
endogenous or nonlinear models, we use a classical exogenous
example in order to provide a contrast with a clear undisputed
benchmark -- the standard linear quantile regression. The
benchmark provides a reliable and high-quality estimation method
for the exogenous model. In this regard, the performance of the LT
estimation and inference, reported in Table \ref{table2} and Table
\ref{table3}, is encouraging.

Table \ref{table2} summarizes the performance of  LTE's and the
standard quantile regression estimator. Table \ref{table3}
compares the performance of the LT confidence intervals to the
standard inference method for quantile regression implemented in
S-plus 4.0. The reported results are averaged across the slope
parameters. The root mean square errors of the LTE's are no larger
than those of quantile regression. Other criteria demonstrate
similar performance of two methods, as predicted by the asymptotic
theory. The coverage of Quasi-posterior quantile confidence
intervals is also close to the nominal level of $90 \%$ in both
small and large samples. It is also noteworthy that the intervals
do not require nonparametric density estimation, as the standard
method requires.

\begin{center}\textbf{[Tables 2 and 3 go here.]}\end{center}

\section{An Illustrative Empirical Application}\label{empiric}

The following illustrates the use of LT estimation in practice.
We consider the problem of forecasting the conditional quantiles
or value-at-risk (VaR) of the Occidental Petroleum (NYSE:OXY)
security returns. The problem of forecasting quantiles of return
distributions is not only important for economic analysis, but is
fundamental to the real-life activities of financial firms. We
offer an econometric analysis of a dynamic conditional quantile
forecasting model, and show that the LTE approach provides a
simple and effective method of estimating such models (despite
the difficulties inherent in the estimation).

The dataset consists of 2527 daily observations (September, 1986
-- November, 1998) on
\begin{itemize}
\item[] $Y_t$, the one-day returns of the Occidental Petroleum
(NYSE:OXY) security, \item[] $X_t$, a vector of returns and prices
of other securities that affect the distribution of $Y_t$: a
constant, lagged one-day return of Dow Jones Industrials (DJI),
the lagged return on the spot price of oil (NCL, front-month
contract on crude oil on NYMEX), and the lagged return $Y_{t-1}$.
\end{itemize}

The choice of variables follows a general principle in which the
relevant conditioning information for estimating value-at-risk of
a stock return, $X_t$, may contain such variables as a market
index of corresponding capitalization and type (for instance, the
$S\&P 500$ returns for a large-cap value stock), the industry
index, a price of a commodity or some other traded risk that the
firm is exposed to, and lagged values of its stock price.

Two functional forms of predictive $\tau$-th quantile regressions
were estimated: \bs \ba{lll}
 & \text{ Linear Model : } \qquad  &  \qquad Q_{\sss Y_{t+1}}(\tau|I_t, \theta(\tau)) = X_t' \; \theta(\tau),  \\
 & \text{ Dynamic Model:  } \qquad & \qquad
Q_{\sss Y_{t+1}}(\tau|I_t, \theta(\tau), \varrho(\tau)) =
X_t'\theta(\tau) + \varrho(\tau)\cdot Q_{\sss Y_{t}}(\tau|I_{t-1},
\theta(\tau), \varrho(\tau)), \ea
\end{split}\end{align}
where $Q_{\sss Y_{t+1}}(\tau|I_t, \theta(\tau))$ denotes the
$\tau$-th conditional quantile of $Y_{t+1}$ conditional on the
information $I_t$ available at time $t$. In other words, $Q_{\sss
Y_{t+1}}(\tau|I_t, \theta(\tau))$ is the value-at-risk at the
probability level $\tau$. The idea behind the dynamic models is
to better incorporate the entire past information and better
predict  risk clustering, as introduced by \cite{engle:caviar}.
The nonlinear dynamic models described by \cite{engle:caviar} are
appealing, but appear to be difficult to estimate using
conventional extremum methods, see \cite{engle:caviar} for
discussion. An extended empirical analysis of the linear model is
given in \cite{vchern2}.

The LT estimation and inference strategy is based on
 the \cite{koenker:1978} criterion function, \bsnumber\label{KB}
L_n(\theta, \varrho) = - \sum_{t=s}^n
    w_t(\tau) \rho_\tau(Y_t - Q_{\sss Y_{t}}(\tau|I_{t-1},
    \theta, \varrho)),
\end{split}\end{align}
where $\rho_{\tau}(u) = (\tau -1(u<0))u$. This criterion function
is similar to that described in Example 1, with the exception that
there is no censoring.  The starting value $s =100$ initializes
the recursive specification so that the imputed initial
conditions (taken to be the marginal quantiles) have a
numerically negligible effect.

In the first step, we constructed the LT estimates using the flat
weights
$$
w_t(\tau) =1/\tau(1- \tau) \text{ for each } \ t=s,...,T.
$$
The results of the first step are not presented here, but they are
very similar to those reported below. Because the weights are not
optimal, the information equality does not hold, hence
Quasi-posterior quantiles are not valid for confidence intervals.
However, the confidence intervals suggested in Theorem 4 lead to
asymptotically valid inference. Under the assumption of correct
dynamic specification, stationary sampling, and the conditions
specified in Proposition 3, the LTE's are consistent and
asymptotically normal
 \bsnumber\label{var_covar}
( \widehat \varrho(\tau),  \widehat \theta(\tau)' )' \ind
\mathcal{N} \(0, J(\theta_0)^{-1} \Omega(\theta_0)
J(\theta_0)^{-1} \),
 \end{split}\end{align}
where for $\nabla q_t(\tau) = \partial Q_{\sss
Y_{t}}(\tau|I_{t-1}, \varrho(\tau),    \theta(\tau))/ \partial (
\varrho , \theta')'$  and $q_t(\tau) = Q_{\sss
Y_{t}}(\tau|I_{t-1}, \varrho(\tau), \theta(\tau))$,
$$
J(\theta_0)= E f_{\sss Y_{t}|I_{t-1}}(q_t(\tau) ) \nabla
q_t(\tau) \nabla q_t(\tau)',
$$
and for $\frac{\Delta_n(\theta_0)}{\sqrt{T-s}} =
\frac{1}{\sqrt{T-s}} \sum_{t=s}^T \[\tau - 1(Y_t < q_t(\tau))\]
\nabla q_t(\tau)$,
$$
\Omega(\theta_0) = \lim_{\sss T \to \infty} \frac{1}{T-s} E
\Delta_n(\theta_0) \Delta_n(\theta_0)' = \tau(1-\tau) E \nabla
q_t(\tau) \nabla q_t(\tau)'.
$$
If the model is not correctly specified, then, for example, the
\cite{newey_west} estimator provides a consistent and robust
procedure for estimation of the limit variance $\Omega(\theta_0)$.

The estimation of the matrix $J(\theta_0)^{-1}$ can be done
through the use of nonparametric methods as in \cite{powell_lad}.
Alternatively, as suggested in Theorem 4, we can use the
variance-covariance matrix of the MCMC sequence of parameter
draws multiplied by $n=(T-s)$ as a consistent estimate of
$J(\theta_0)^{-1}$. Plugging the estimates into the variance
expression (\ref{var_covar}), we obtain the standard errors and
confidence intervals that are qualitatively similar to those
reported in Figures 4-7.

In order to illustrate the use of Quasi-posterior quantiles
(Theorem 3) and improve estimation efficiency, we also carried out
the second step estimation using the Koenker-Bassett criterion
function (\ref{KB}) with the weights $$ \widehat w_t(\tau) =
\frac{h}{[ Q_{\sss Y_{t}}(\tau+h/2|I_{t-1}, \widehat
\varrho(\tau), \widehat \theta(\tau)) - Q_{\sss Y_{t}}(\tau- h/2
|I_{t-1}, \widehat \varrho(\tau), \widehat \theta(\tau)) ]} \cdot
\frac{1}{\tau(1-\tau)},$$ where $h \propto C n^{-1/3}$ and $C>0$
is chosen using the rule given in \cite{quantint}. Under the
assumption of correct dynamic specification, these weights imply
the generalized information equality, which validates
Quasi-posterior quantiles for inference purposes, as in
(\ref{np_weights})-(\ref{np_equality}). The following analysis is
based on the second step estimates. The $.05$-th, $.5$-th, and
$.95$-th Quasi-posterior quantiles are computed for each
coefficient $\theta_j(\tau)$ $(j=1,...,4)$ and $\varrho(\tau)$,
and then used to form the point estimates and the
$90\%$-confidence intervals, which are reported  Figures 4-7 for
 $\tau = .2, .4,...,.8$.

Figures 2 and 3 present the estimated surfaces of the conditional
VaR functions of the dynamic model and linear models,
respectively, plotted in the time-probability level coordinates,
$(t, p),$ ($ p=\tau$ is the quantile index.)  We report VaR for
many values of $\tau$. The conventional VaR reporting typically
involves the probability levels at a given $\tau$. Clearly, the
whole  VaR surface formed by varying $\tau$ represents a more
 complete depiction of conditional risk.

The dynamics depicted in Figures 2 and 3 unambiguously indicate
certain dates on which market risk tends to be much  higher than
its usual
 level. The difference between the linear and the recursive model is
 also
striking. The risk surface generated by the recursive model is
much smoother and is much more persistent. Furthermore,  this
difference is statistically significant, as Figure 7 shows.

Focusing on the recursive model, let us examine the economic and
statistical interpretation of the slope coefficients $\widehat
\theta_2(\cdot)$, $\widehat \theta_3(\cdot)$, $\widehat
\theta_4(\cdot)$, $\widehat \varrho\(\cdot\)$, plotted in Figures
4-7.

The coefficient on the lagged oil price return,   $\widehat
\theta_2(\cdot)$, is insignificantly positive  in the left and
right tails of the conditional return
 distribution.  It is
 insignificantly negative in the middle part.
 The coefficient on the lagged DJI return, $\widehat \theta_3(\cdot)$, in contrast, is significantly positive for all
values of
 $\tau$. We
 also notice a sharp increase in the middle range.
 Thus,  in addition to
 the  strong positive relation between the individual
 stock return and the market  return (DJI) (dictated by the fact that $\theta_2(\cdot) > 0$ on $(0.2,0.8)$) there is also
  additional sensitivity of the median of the security return to the market movements.

The coefficient on the own lagged return, $\widehat
\theta_4(\cdot)$, on the other hand, is significantly negative,
except
 for values of $\tau$ close to $0$. This may be interpreted as a  reversion effect in the central part of the distribution. However,
the lagged return does not appear to
 significantly shift the quantile function in the tails. Thus, the lagged return
 is more  important for the determination of
 intermediate risks.

Most importantly, the dynamic coefficient $\widehat
\varrho\(\cdot\)$ on the lagged VaR is significantly negative in
the low quantiles and in the high quantiles, but is insignificant
in the middle range. The significance of $\widehat
\varrho\(\cdot\)$ is a strong evidence in favor  of the recursive
specification. The magnitude and sign of $\widehat
\varrho\(\cdot\)$ indicates both the reversion and significant
risk clustering effects in the tails of the distribution (see
Figure 7). As expected, there is zero effect over the middle
range, which is consistent with the random walk properties of the
stock price. Thus, the dynamic effect of lagged VaR is much more
important for the tails of the quantile function, that is for
risk management purposes.

\vspace{-.1in}
\section{Conclusion}\label{conclusion}

In this paper, we study the  Laplace-type Estimators or
Quasi-Bayesian Estimators  that we define using common
statistical, non-likelihood based criterion functions. Under mild
regularity conditions these estimators are $\sqrt{n}$-consistent
and asymptotically normal, and Quasi-posterior quantiles provide
asymptotically valid confidence intervals. A simulation study and
an empirical example illustrate the properties of the proposed
estimation and inference methods. These results show that in many
important cases the Quasi-Bayesian estimators provide useful
alternatives to the usual extremum estimators. In ongoing work,
we are extending the results to models in which
$\sqrt{n}$-convergence rate and asymptotic normality do not hold,
including the maximum score problem.

\section*{ Acknowledgments} We thank the editor for the invitation of this
paper to Journal of Econometrics and an anonymous referee for
prompt and highest quality feedback. We thank Xiahong Chen, Shawn
Cole, Gary Chamberlain, Ivan Fernandez, Ronald Gallant, Jinyong
Hahn, Bruce Hansen, Chris Hansen, Jerry Hausman, James Heckman, Bo
Honore, Guido Imbens, Roger Koenker, Shakeeb Khan, Sergei Morozov,
Whitney Newey, Ziad Nejmeldeen, Stavros Panageas, Chris Sims,
George Tauchen, and seminar participants at Brown University,
Duke-UNC Triangle Seminar, MIT, MIT-Harvard, University of
Chicago, Princeton University, University of Wisconsin at
Madison, University of Michigan, Michigan State University,
Texas-AM University,  the Winter meeting of the Econometric
Society, the 2002 European Econometric Society Meeting in Venice
for insightful comments. We gratefully acknowledge the financial
support provided by the U.S. National Science Foundation grants
SES-0214047 and  SES-0079495. %\baselineskip=\baselineskip \small

%\linespread{1.15}
\appendix 

\section{Appendix of Proofs}

\subsection{Proof of Theorem \ref{theorem2}}

It suffices to show \bsnumber\label{density total convergence}
 \int_{H_n} | h |^\alpha \Big | p_n^*(h)
- p^*_{\infty}(h)  \Big | dh \inp 0
\end{split}\end{align}
for all $\alpha \geq 0$. Our arguments follow those in
\cite{bickel2}  and \cite{ibragimov}, as presented by
\cite{lehmann}. As indicated in the text, the main difference are
in part 2, and are due to (i) the non-likelihood setting, (ii)
the use of Huber-like conditions in Assumption \ref{second order
expansion} to handle discontinuous criterion functions, (iii)
allowing more general loss functions, which are needed for
construction of confidence intervals.

Throughout this proof the range of integration for $h$ is
implicitly understood to be $H_n$.  For clarity of the argument,
we limit exposition only to the case where $J_n(\theta_0)$ and
$\Omega_n(\theta_0)$ do not depend on $n$. The more general case
follows similarly.

\noindent\textbf{Part 1.} Define \bsnumber\label{define h} h
\equiv \sqrt{n}\(\theta - T_n\), \ T_n = \theta_0 + \frac1n
J\(\theta_0\)^{-1} \Delta_n\(\theta_0\), \ U_n \equiv
\frac{1}{\sqrt{n}} J(\theta_0)^{-1}
\Delta_n(\theta_0),\end{split}\end{align} then \bs p_n^*\(h\)
\equiv &
\frac{1}{\sqrt{n}}p_n\(h/\sqrt{n}+\theta_0+U_n/\sqrt{n}\) \\
= &
 \frac{
\pi\(\frac{h}{\sqrt{n}}+ T_n \)
\exp\(L_n\(\frac{h}{\sqrt{n}}+T_n\)\) } { \int_{H_n}
\pi\(\frac{h}{\sqrt{n}}+ T_n \)
\exp\(L_n\(\frac{h}{\sqrt{n}}+T_n\)\) dh
}\\
= & \frac{ \pi\(\frac{h}{\sqrt{n}}+ T_n \) \exp\( \omega(h)\) } {
\int_{H_n} \pi\(\frac{h}{\sqrt{n}}+ T_n \) \exp\( \omega(h) \)
dh}  \\ = & \frac{ \pi\(\frac{h}{\sqrt{n}}+ T_n \) \exp\(
\omega(h)\) } { C_n },
\end{split}\end{align}
where  \bsnumber\label{define wh} \omega\(h\) \equiv L_n\(T_n +
\frac{h}{\sqrt{n}}\) -L\(\theta_0\) - \frac{1}{2n}
\Delta_n\(\theta_0\)' J\(\theta_0\)^{-1} \Delta_n\(\theta_0\),
\end{split}\end{align}
and
$$
C_n \equiv \int_{H_n} \pi\(\frac{h}{\sqrt{n}}+ T_n \) \exp\(
\omega(h) \) dh.
$$

\vspace{.1in}

\noindent Part 2 shows that for each $\alpha \geq 0$,
\bsnumber\label{pre density total convergence} A_{1n} \equiv
\int_{H_n} \vert h\vert^\alpha \bigg\vert \exp\(\omega(h)\) &\pi\(T_n
+ \frac{h}{\sqrt{n}}\) - \exp\(-\frac12 h'J\(\theta_0\) h\)
\pi\(\theta_0\) \bigg\vert dh \overset{p}{\longrightarrow} 0.
\end{split}\end{align}
Given (\ref{pre density total convergence}), taking $\alpha =0$
we have
 \bsnumber\label{constant
convergence} C_n  \inp \int_{\Bbb{R}^d} e^{-\frac12
h'J\(\theta_0\) h} \pi\(\theta_0\) dh  = \pi\(\theta_0\)
(2\pi)^{\frac{d}{2}} |\det J\(\theta_0\)|^{-1/2},
\end{split}\end{align}
hence
$$
C_n = O_p(1).
$$

Next note
$$
\text{ left side of (\ref{density total convergence})} \ \equiv
\int_{H_n} | h |^\alpha \Big | p_n^*(h) - p^*_{\infty}(h) \Big |
dh  = A_n\cdot C_n^{-1},
$$
where \bs A_n \equiv \int_{H_n} \vert h\vert^\alpha \bigg\vert
e^{\omega(h)} & \pi\(T_n + \frac{h}{\sqrt{n}}\)- (2\pi)^{-d/2} |\det
J\(\theta_0\)|^{1/2} \exp \(-\frac{1}{2} h'J\(\theta_0\)h\)\cdot
C_n \bigg\vert dh.
\end{split}\end{align}
Using (\ref{constant convergence}), to show (\ref{density total
convergence}) it suffices to show that
$A_n\overset{p}{\longrightarrow} 0$. But $$A_n \leq A_{1n} +
A_{2n}$$ where $$A_{2n} \equiv \int_{H_n} \vert h\vert^\alpha
\bigg\vert C_n (2\pi)^{-d/2}|\det J\(\theta_0\)|^{1/2}
\exp\(-\frac{1}{2} h'J\(\theta_0\)h\) - \pi\(\theta_0\)
\exp\(-\frac12 h'J\(\theta_0\) h\) \bigg\vert dh.$$ Then by
(\ref{pre density total convergence})
$$
A_{1n} \overset{p}{\longrightarrow} 0,$$
 and   \bs A_{2n}
& = \bigg\vert C_n (2\pi)^{-d/2}| \det J\(\theta_0\)|^{1/2}
-\pi\(\theta_0\) \bigg\vert \int_{H_n} \vert h\vert^\alpha
\exp\(-\frac12 h'J\(\theta_0\) h\) dh \\
& \overset{p}{\longrightarrow} 0.
\end{split}\end{align}

\noindent\textbf{Part 2.} It remains only to show (\ref{pre
density total convergence}). Given Assumption \ref{second order
expansion} and definitions in (\ref{define h}) and (\ref{define
wh}),  write \bs \omega\(h\) & = \Delta_n(\theta_0) ' \( \frac{U_n
+h }{\sqrt{n}}  \) - \frac{1}{2} n
 \( \frac{U_n + h}{\sqrt{n}} \)' J(\theta_0) \(
 \frac{U_n + h}{\sqrt{n}}  \) \\ & - \frac{1}{2n}
\Delta_n\(\theta_0\)' J\(\theta_0\)^{-1} \Delta_n\(\theta_0\) +    R_n\( \frac{h}{\sqrt{n}} + T_n  \)   \\
& = -\frac12 h'J\(\theta_0\)h +
 R_n\( \frac{h}{\sqrt{n}} + T_n \).
\end{split}\end{align}
Split the integral $A_{1n}$ in (\ref{pre density total
convergence}) over three separate areas:
\begin{itemize}
\item Area (i) \ \  : $\vert h \vert \leq M$,
\item Area (ii)\ \ : $M \leq \vert h\vert \leq \delta
\sqrt{n}$,
\item Area (iii)\ : $\vert h\vert \geq
\delta \sqrt{n}$.
\end{itemize}

\noindent Each of these areas
is implicitly understood to intersect with the range
of integration for $h$, which is $H_n$.

\noindent \textbf{Area (i):} We will show that for each $0 < M <
\infty$ and each $\epsilon >0$ \bsnumber\label{area1} \liminf_n
P_* \Bigg \{ \int_{\vert h \vert \leq M} \vert h\vert^\alpha &
\bigg\vert \exp\(\omega(h)\) \pi\(T_n + \frac{h}{\sqrt{n}}\) \\ &-
\exp\(-\frac12 h'J\(\theta_0\) h\) \pi\(\theta_0\) \bigg\vert dh
< \epsilon \Bigg \} \geq 1- \epsilon.
\end{split}\end{align}
This is proved by showing that \bsnumber\label{within M
convergence} \sup_{\vert h\vert \leq M}\vert  h\vert^\alpha
\bigg\vert \exp\(\omega(h)\) & \pi\(T_n + \frac{h}{\sqrt{n}}\) -
\exp\(-\frac12 h'J\(\theta_0\) h\) \pi\(\theta_0\) \bigg\vert
\overset{p}{\longrightarrow} 0.
\end{split}\end{align}
Using the definition of $\omega\(h\)$, (\ref{within M
convergence}) follows from: \bs &(a) \ \sup_{\vert h \vert \leq
M} \bigg\vert \pi\(\frac{h}{\sqrt{n}} + T_n\) - \pi\(\theta_0\)
\bigg\vert \overset{p}{\longrightarrow} 0,\ \ \  \ (b) \
\sup_{\vert h \vert \leq M} \bigg\vert R_n\(\frac{h}{\sqrt{n}} +
T_n\) \bigg\vert \overset{p}{\longrightarrow} 0,
\end{split}\end{align}
where (a) follows from the continuity of $\pi\(\cdot\)$ and
because by Assumption 4.ii-4.iii: \bsnumber\label{O p 1}
\frac{1}{\sqrt{n}} J\(\theta_0\)^{-1} \Delta_n\(\theta_0\)  =
O_p\(1\).
\end{split}\end{align}
Given (\ref{O p 1}), (b) follows from Assumption 4.iv, since
 \bs
\sup_{\vert h \vert \leq M} \bigg\vert T_n + \frac{h}{\sqrt{n}}
-\theta_0 \bigg\vert = O_p(1/\sqrt{n}).
\end{split}\end{align}

\noindent\textbf{Area (ii)}: We show that for each $\epsilon
> 0$ there exist large $M$ and small $\delta>0$ such that
\bsnumber\label{area2} \liminf_n P_* \Bigg \{\int_{M < \vert h
\vert < \delta\sqrt{n}} &\vert h\vert^\alpha \bigg\vert
\exp\(\omega(h)\) \pi\(T_n + \frac{h}{\sqrt{n}}\) \\ &-
\exp\(-\frac12 h'J\(\theta_0\) h\) \pi\(\theta_0\) \bigg\vert dh
< \epsilon \Bigg \} \geq 1-\epsilon.
\end{split}\end{align}
Since the integral of the second term is finite and can be made
arbitrarily small by setting $M$ large, it suffices to show that
for each $\epsilon
> 0$ there exist large $M$ and small $\delta>0$ such that
\bsnumber\label{area21} \liminf_n P_* \Bigg \{\int_{M < \vert h
\vert < \delta\sqrt{n}} &\vert h\vert^\alpha \bigg\vert
\exp\(\omega(h)\) \pi\(T_n + \frac{h}{\sqrt{n}}\)  \bigg\vert dh <
\epsilon \Bigg \} \geq 1-\epsilon.
\end{split}\end{align}
In order to do so, it suffices to show that for sufficiently
large $M$ as $n \rightarrow \infty$
\bsnumber\label{area2_suffices} \exp\(\omega(h)\) \pi\(T_n +
\frac{h}{\sqrt{n}}\) \leq C \exp\(-\frac14 h'J\(\theta_0\) h\),
\text{ for all } M < \vert h \vert < \delta\sqrt{n}.
\end{split}\end{align}
By assumption  $\pi\(\cdot\) < K$, so we can drop it from
consideration.

\noindent By definition of $\omega(h)$  \bs \exp\(\omega(h)\) \leq
\exp \Bigg ( &-\frac12 h'J\(\theta_0\)h  + \bigg\vert R_n\(T_n
+\frac{h}{\sqrt{n}}\)\bigg\vert  \Bigg).
\end{split}\end{align}
Since $|T_n - \theta_0|= o_p(1)$, for any $\delta>0$ wp
$\rightarrow 1$ \bs \bigg\vert T_n + \frac{h}{\sqrt{n}} -
\theta_0 \bigg\vert < 2\delta, \quad\text{for all}\quad \vert
h\vert \leq \delta\sqrt{n}.
\end{split}\end{align}
Thus, by  Assumption 4.\textbf{iv}(a) there exists some small
$\delta$ and large $M$ such that \bs \lim\inf_n P_*\Bigg \{\sup_{
M \leq \vert h\vert \leq \delta\sqrt{n}} \frac{|R_n\(T_n +
\frac{h}{\sqrt{n}}\)|}{|h +
\frac{1}{\sqrt{n}}J\(\theta_0\)^{-1}\Delta_n\(\theta_0\) |^2}
 \leq \frac14 \text{mineig} \(J\(\theta_0\)\)  \Bigg \}
\geq 1-\epsilon.
\end{split}\end{align}
Since $\frac1n \big\vert J\(\theta_0\)^{-1}\Delta_n\(\theta_0\)
\big\vert^2 = O_p\(1\)$, for some $C > 0$ \bsnumber\label{exp
bound} &\liminf_n P_*\Bigg \{\exp\(\omega(h)\) \leq C \exp\(-\frac14
h'J\(\theta_0\) h\)\Bigg \} \\
&\geq   \liminf_n P_*\Bigg \{e^{\omega(h)} \leq C \exp\(-\frac12
h'J\(\theta_0\) h+\frac14 \text{mineig}\(J\(\theta_0\)\) |h|^2\)
\Bigg \} \\ & \geq 1- \epsilon.
\end{split}\end{align}
(\ref{exp bound}) implies (\ref{area2_suffices}), which in turn implies
(\ref{area2}).

\noindent \textbf{Area (iii):} We will show that for each
$\epsilon
> 0$ and each $\delta > 0$, \bsnumber\label{area3} \liminf_n P_*
\Bigg \{\int_{ h \geq \delta\sqrt{n}} &\vert h\vert^\alpha
\bigg\vert \exp\(\omega(h)\) \pi\(T_n + \frac{h}{\sqrt{n}}\) \\ &-
\exp\(-\frac12 h'J\(\theta_0\) h\) \pi\(\theta_0\) \bigg\vert dh
< \epsilon \Bigg \} \geq 1-\epsilon.
\end{split}\end{align}
The integral of the second term clearly goes to $0$ as
$n\rightarrow\infty$. Therefore we only need to show
 \bs &\int_{\vert h\vert \geq \delta\sqrt{n}}
\vert h\vert^\alpha e^{\omega\(h\)} \pi\(T_n +\frac{h}{\sqrt{n}}\)
dh \inp 0.
\end{split}\end{align}
Recalling the definition of $h$, the term is bounded by \bs
\sqrt{n}^{\alpha+1} \int_{\vert \theta - T_n \vert \geq \delta}
\bigg\vert T_n - \theta\bigg\vert^{\alpha} \pi\(\theta\) &
\exp\Bigg( L_n\(\theta\)-L_n\(\theta_0\)  - \frac{1}{2n}
\Delta_n\(\theta_0\)' J\(\theta_0\)^{-1} \Delta_n\(\theta_0\)
\Bigg ) d\theta.
\end{split}\end{align}
Since $T_n - \theta_0 \overset{p}{\longrightarrow} 0$, wp
$\rightarrow 1$ this is bounded by
 \bs
K_n \cdot C \cdot \sqrt{n}^{\alpha+1} \int_{\vert \theta -
\theta_0\vert \geq
 \delta/2}\ \(1+\big\vert \theta\big\vert^{\alpha}\) \pi\(\theta\)
\exp\( L_n\(\theta\)-L_n\(\theta_0\) \) d\theta,
\end{split}\end{align}
where $$K_n = \exp\(- \frac{1}{2n} \Delta_n\(\theta_0\)'
J\(\theta_0\)^{-1} \Delta_n\(\theta_0\)\) = O_p(1).$$

\noindent By Assumption \ref{consistency assumption} there exists
$\varepsilon
> 0$ such that \bs \liminf_{n \rightarrow \infty} P_*\Bigg \{ \sup_{\vert \theta-\theta_0  \vert \geq \delta/2}
e^{L_n\(\theta\) - L_n\(\theta_0\)} \leq e^{-n\varepsilon}\Bigg \}
= 1
\end{split}\end{align}
Thus, wp $\rightarrow 1$ the entire term is bounded by
 \bsnumber\label{conclude c}
K_n \cdot C\cdot \sqrt{n}^{\alpha+1} \cdot e^{-n\epsilon}
\int_{\Theta} \vert \theta \vert^\alpha \pi\(\theta\)d\theta
=o_p(1).
\end{split}\end{align}
Here observe that compactness is only used to insure that
 \bsnumber\label{drop compactness}
\int_{\Theta} \vert \theta \vert^\alpha \pi\(\theta\)d\theta<
\infty. \end{split}\end{align} Hence by replacing compactness
with the condition (\ref{drop compactness}), the conclusion
(\ref{conclude c}) is not affected for the given $\alpha$.

 The entire proof is now completed by combining (\ref{area1}),
(\ref{area2}), and (\ref{area3}). \hfill $\blacksquare$

\subsection{Proof of Theorem \ref{theorem3}} For
clarity of the argument, we limit exposition only to the case
where $J_n(\theta_0)$ and $\Omega_n(\theta_0)$ do not depend on $n$. The more general case follows similarly.
\footnote{Indeed, given Assumption 4, we can split $\{1,2,..,\}$ into subsequences along which $J_n(\theta_0)$ and $\Omega_n(\theta_0)$ converge to some fixed matrices (depending on the subsequence), and work with each subsequence to obtain required approximations, $Z_n = \xi + U_n$, where $\xi$ and $U_n$ are defined in terms of the fixed matrices. Then, along each subsequence, we can replace $\xi$ and $U_n$ by versions that depend on $J_n(\theta_0)$ and $\Omega_n(\theta_0)$, incurring an $o(1)$ error.}

Recall that \bs h = \sqrt{n}\(\theta - \theta_0\) -
J\(\theta_0\)^{-1} \Delta_n\(\theta_0\)/\sqrt{n}.
\end{split}\end{align}
Define $U_n= J(\theta_0)^{-1} \Delta_n(\theta_0)/\sqrt{n}$.
Consider the objective function
$$
Q_{n} (z) = \int_{H_n} \rho( z -h -U_n) p_n^*(h) dh,
$$
which is minimized at $\sqrt{n} (\widehat \theta - \theta_0)$.
Also define
$$
Q_{\infty} (z) = \int_{\Bbb{R}^d} \rho(z -h-U_n) p_{\infty}^*(h)
dh.
$$
which is minimized at a random vector denoted $Z_n$. Define
$$
\xi = \arg\inf_{z \in \Bbb{R}^d} \Bigg \{ \int_{\Bbb{R}^d}
\rho(z-h) \ p_{\infty}^*(h) \ dh \Bigg \}.
$$
Note that solution is unique and finite by Assumption \ref{loss
function} parts (ii) and (iii) on the loss function $\rho$. When
$\rho$ is symmetric, $\xi= 0$ by Anderson's lemma.

\noindent Therefore, $Z_n = \arg\inf_{z\in \Bbb{R}^d}
Q_{\infty}(z)$ equals
$$
Z_n = \xi + U_n = O_p(1).
$$
\noindent Next, we have for any  fixed $z$
$$
 Q_n(z)- Q_{\infty}(z) \inp 0
$$
since by Assumption \ref{loss function}.\textbf{ii} $\rho(h) \leq
1 + |h|^p$  and by $|a+b|^p \leq 2^{p-1} |a|^p + 2^{p-1} |b|^p$
for $p \geq 1$:
 \bs
|Q_n(z) - Q_{\infty}(z)| &\leq \int_{H_n} \(1 + |z-h-U_n|^p\)
\vert
p^*_n(h) - p^*_{\infty}(h) \vert dh  \\
& + \hspace{.5in}\int_{H_n^c} \(1 + |z-h-U_n|^p\) \( p^*_{\infty}(h) \) dh \\
 & \leq  \int_{H_n} \(1 + 2^{p-1} |h|^p + 2^{p-1} |z-U_n|^p\) \vert
p^*_n(h) - p^*_{\infty}(h) \vert dh  \\
&+ \hspace{.5in} \int_{H_n^c} \(1 + 2^{p-1} |h|^p + 2^{p-1}
|z-U_n|^p\) \(
p^*_{\infty}(h) \) dh \\
& \leq  \int_{H_n} \(1 + 2^{p-1} |h|^p + O_p(1) \) \(
p^*_n(h) - p^*_{\infty}(h) \) dh\\
&+ \hspace{.5in} \int_{H_n^c} \(1 + 2^{p-1} |h|^p + O_p(1)\) \(
p^*_{\infty}(h) \) dh  = o_p(1),
\end{split}\end{align}
where $o_p(1)$-conclusion is by Theorem
\ref{theorem2} and exponentially small
tails of the normal density (Gaussian measure of $H_n^c$ converges
to zero).

\noindent Now note that $Q_{n}(z)$ and $Q_{\infty}(z)$ are convex
and finite, and $Z_n= \arg\inf_{z\in \Bbb{R}^d} Q_{\infty}(z)=
O_p(1)$. By the convexity lemma of \cite{pollard1}, pointwise
convergence entails the uniform convergence over compact sets $K$:
$$
\sup_{z \in K} \Big | Q_{n}(z) - Q_{\infty}(z) \Big | \inp 0.
$$
Since $Z_n=O_p(1)$, uniform convergence and convexity arguments
like those in \cite{jureckova:convex} imply that $\sqrt{n}
(\widehat \theta - \theta_0) - Z_n \inp 0$, as shown below.
\hfill $\blacksquare$

\vspace{.1in}
\noindent \textbf{Proof of } $Z_n - \sqrt{n} (\widehat \theta -
\theta_0)=o_p(1)$. The proof follows by extending slightly the
convexity argument of \cite{jureckova:convex} and \cite{pollard1}
to the present context. Consider a ball $B_{\delta}(Z_n)$ with
radius $\delta>0$, centered at $Z_n$, and let $z =Z_n +  d v$,
where $v$ is a unit direction vector such that $|v| = 1$ and $d
> \delta$ is a scalar. Because $Z_n = O_p(1)$, for any $\delta>0$ and
$\epsilon>0$, there exists $K>0$ such that
$$
 \liminf_{n \rightarrow \infty} P_* \Big \{ E_n = \{ B_{\delta}(Z_n)\in
B_K(0)\} \Big \}\geq 1- \epsilon.
$$

 By convexity, for any $z=Z_n +  d v$ constructed so, it follows
 that
 \bsnumber\label{star2} \frac{\delta}{d} \big ( Q_{n} (z)
-  Q_{n} (Z_n) \big) \geq Q_{n} (z^*) - Q_{n} (Z_n),
\end{split}\end{align}
 where
$z^*$ is a point of boundary of $B_{\delta}(Z_n)$ on the line
connecting $z$ and $Z_n$. By the uniform convergence of $Q_n(z)$
to $Q_{\infty}(z)$ over any compact set $B_K(0)$, whenever $E_n$
occurs: \bs \frac{\delta}{d} \big ( Q_{n} (z) - Q_{n} (Z_n) \big)
&\geq Q_{n}
(z^*) - Q_{n} (Z_n) \\
& \geq Q_{\infty} (z^*) - Q_{\infty} (Z_n) + o_p(1)  \geq V_n
+o_p(1),
\end{split}\end{align}
where $V_n>0$ is a uniformly in $n$ positive variable, because
$Z_n$ is the unique optimizer of  $Q_{\infty}$. That is, there
exists an $\eta>0$ such that $\liminf_n P(V_n> \eta) \geq 1-
\epsilon$. Hence we have with probability at least as big as $1-
3\epsilon$ for large $n$: \bs \frac{\delta}{d} \big ( Q_{n} (z)
-  Q_{n} (Z_n) \big) > \eta.
\end{split}\end{align}
Thus, $\sqrt{n}(\widehat \theta - \theta_0)$ eventually belongs
to a complement of $B_{\delta}(Z_n)$  with probability at most $3
\epsilon$. Since we can set  $\epsilon$ as small as we like by
picking (a) sufficiently large $K$, and (b) sufficiently large
$n$, and (c) sufficiently small $\eta>0$,
 it follows that
$$\limsup_{n \rightarrow \infty} P^* \Big \{ |Z_n-  \sqrt{n}(\widehat \theta -
\theta_0)| > \delta \Big \} = 0.
$$
Since this is true for any $\delta>0$, it follows that
$$
Z_n-  \sqrt{n}(\widehat \theta - \theta_0)  = o_p(1). \ \
\blacksquare
$$

\subsection{Proof of Theorem 3} For clarity of the argument, we
limit exposition only to the case where $J_n(\theta_0)$ and
$\Omega_n(\theta_0)$ do not depend on $n$. The more general case
follows similarly. We defined
$$
F_{g,n}( x) = \int_{\theta \in \Theta: g(\theta) \leq x } p_n
(\theta) d \theta.
$$
Evaluate it at $x=g(\theta_0) + s/\sqrt{n}$ and change the
variable of integration
$$
H_{g,n}( s) =F_{g,n}(g(\theta_0) + s/\sqrt{n})= \int_{h \in H_n :
g(\theta_0 + h/\sqrt{n} + U_n/\sqrt{n}) \leq  g(\theta_0 ) +
s/\sqrt{n} } p^*_n (h) d h.
$$
Define also
$$
\widehat H_{g,n}( s) = \int_{h \in \Bbb{R}^d : g(\theta_0 +
h/\sqrt{n} + U_n/\sqrt{n}) \leq  g(\theta_0 ) + s/\sqrt{n} }
p^*_{\infty} (h) d h
$$
and
$$
H_{g,\infty}( s) = \int_{h \in \Bbb{R}^d: \nabla g(\theta_0)'(
h/\sqrt{n} + U_n/\sqrt{n}) \leq s/\sqrt{n} } p^*_{\infty} (h) d h.
$$
By definition of total variation of moments norm and Theorem 1
$$
\sup_{s} | H_{g,n}( s) - \widehat H_{g,n}( s)|  \inp 0,
$$
where the $\sup$ is taken over the support of $H_{g,n}(s)$.

By the uniform continuity of the integral of the normal density
with respect to the boundary of integration
$$
\sup_{s} | \widehat H_{g,n}( s) - H_{g,\infty}( s)| \inp 0,
$$
which implies
$$
\sup_{s} | H_{g,n}( s) -  H_{g,\infty}( s)| \inp 0.
$$
where the $\sup$ is taken over the support of $H_{g,n}(s)$.

The convergence of distribution function implies the convergence
of quantiles at continuity points of distribution functions, see
e.g. \cite{billingsley:1994}, so
$$
H^{-1}_{g,n}(\alpha) - H^{-1}_{g,\infty}(\alpha) \inp 0.
$$
\noindent Next observe
$$
H_{g,\infty}(s) = P\Big \{ \nabla g(\theta_0)' \mathcal{N} ( U_n,
J^{-1}(\theta_0)) < s \Big |U_n \Big \},
$$
so
$$
 H^{-1}_{g,\infty}(\alpha) = \nabla
g(\theta_0)' U_n + q_{\alpha} \sqrt{ \nabla_{\theta}
 g(\theta_0)' J^{-1}(
\theta_0)\nabla_{\theta}g(\theta_0) },
$$
where $q_{\alpha}$ is the $\alpha$-quantile of $\mathcal{N}(0,1)$.

\noindent Recalling that we defined $c_{g,n}(\alpha) =
F_{g,n}^{-1}(\alpha)$, by quantile equivariance with respect to
the monotone transformations $$ H^{-1}_{g,n}(\alpha) =
\sqrt{n}\(c_{g,n}(\alpha) - g(\theta_0)\)$$ so that
$$
\sqrt{n}\(c_{g,n}(\alpha) - g(\theta_0)\) = \nabla g(\theta_0)'
U_n + q_{\alpha} \sqrt{ \nabla_{\theta}
 g(\theta_0)' J^{-1}(
\theta_0)\nabla_{\theta}g(\theta_0) } + o_p(1).
$$
The rest of the result follows by the $\Delta$-method.\hfill
$\blacksquare$

\subsection{Proof of Theorem 4} For clarity of the argument, we
limit exposition only to the case where $J_n(\theta_0)$ and
$\Omega_n(\theta_0)$ do not depend on $n$. In view of Assumption 4, it suffices to show that
 \bsnumber\label{JJ}
\widehat J^{-1}_n(\theta_0)- J^{-1}_n(\theta_0) \inp 0,
 \end{split}\end{align}
and then conclude by the $\Delta$-method.

Recall that \bs h = \sqrt{n}\(\theta - \theta_0\) -
\underbrace{J_n\(\theta_0\)^{-1}
\Delta_n\(\theta_0\)/\sqrt{n}}_{U_n},
\end{split}\end{align}
and the localized Quasi-posterior density for $h$ is \bs
p_n^*\(h\) =
\frac{1}{\sqrt{n}}p_n\(h/\sqrt{n}+\theta_0+U_n/\sqrt{n}\).
\end{split}\end{align}

Note also
 \bs
\widehat J^{-1}_n(\theta_0) & \equiv  \int_{\Theta} n (\theta-
\widehat \theta )(\theta - \widehat \theta )' p_n(\theta) d \theta \\
 & =   \int_{H_n} (h - \sqrt{n} (
\widehat \theta - \theta_0) + U_n ) \cdot (h - \sqrt{n} ( \widehat
\theta - \theta_0) +
U_n  )' p_n^*(h) d  h,  \\
\text{ and }  & \\
J^{-1}_n(\theta_0) &\equiv  \int_{\Bbb{R}^d} h h' p_{\infty}^*(h)
d h.
\end{split}\end{align}

We have, denoting $h = (h_1,...,d_d)$ and  $\widetilde T_n =
(\widetilde T_{n1},...,\widetilde T_{nd})$ where $\widetilde T_n
= \sqrt{n} ( \widehat \theta - \theta_0) - U_n$, for all $i, j
\leq d$
\begin{itemize}
\item[(a)]  $ \int_{H_n}  h_i h_j  \Big ( p_n^*(h)
- p^*_{\infty}(h)  \Big ) dh = o_p(1)$ by Theorem 1, \\
\item[(b)] $\int_{H_n^c}  h_i h_j  \Big ( p^*_{\infty}(h)  \Big ) dh
= o_p(1)$ by definition of $p^*_{\infty}$ and $J_n(\theta_0)$
being uniformly nonsingular, \\
\item[(c)] $\int_{H_n} \underbrace{|\widetilde  T_n|^2}_{= o_p(1)}  \  \Big ( p_n^*(h) - p^*_{\infty}(h)  \Big ) dh   =
o_p(1)$ by Theorem 2, \\
\item[(d)] $\int_{H_n} \underbrace{|\widetilde  T_n|^2}_{= o_p(1)}  \  \Big ( p^*_{\infty}(h)  \Big ) dh   =
o_p(1)$ by Theorem 2, definition of $p^*_{\infty}$, and
$J_n(\theta_0)$ being nonsingular, \\
\item[(e)] $\int_{H_n} h_j \underbrace{\widetilde T_{ni}}_{= o_p(1)}  \  \Big ( p_n^*(h) - p^*_{\infty}(h)  \Big ) dh   =
o_p(1)$ by Theorems 1 and 2, \\
\item[(f)] $\int_{H_n} h_j \underbrace{\widetilde T_{ni}}_{= o_p(1)}  \  \Big ( p^*_{\infty}(h)  \Big ) dh   =
o_p(1)$ by Theorems 1 and 2, definition of $p^*_{\infty}$, and
$J_n(\theta_0)$ being uniformly nonsingular, from which the
required conclusion follows. $\blacksquare$
\end{itemize}

\noindent

\subsection{Proof of Proposition 1} Assumption \ref{consistency assumption} is directly implied by
(\ref{gmm1})-(\ref{gmm3}) and the uniform continuity of $E
m_i\(\theta\)$, as shown in Lemma \ref{amen consistency}. It
remains only to verify Assumption \ref{second order expansion}.

Define the identity
 \bsnumber\label{GMMid}
L_n(\theta) - L_n(\theta_0) & \equiv - \
\underset{\Delta_n(\theta_0)'}{\underbrace { n g_n(\theta_0)'
W(\theta_0) G(\theta_0)}} (\theta - \theta_0) \\
& - \frac12 (\theta - \theta_0)' n
\underset{J(\theta_0)}{\underbrace {G(\theta_0)' W(\theta_0)
G(\theta_0)}}(\theta - \theta_0)  \ \  \ + \ \ \ R_n(\theta).
\end{split}\end{align}

Next, given the definition of $\Delta_n(\theta_0)$ and
$J(\theta_0)$, conditions \textbf{i}, \textbf{ii}, \textbf{iii} of
Assumption 4. are immediate from conditions \textbf{i-iii} of
Proposition 1. Condition \textbf{iv} is verified as follows.
Condition \textbf{iv} of Assumption \ref{second order expansion}
can be succinctly stated as: \bs \text{ for each } \epsilon
> 0  \text{ there exists a } \delta > 0 \ \text{such that}\ \limsup_{n\rightarrow
\infty} P^*\Bigg \{ \sup_{\vert\theta-\theta_0\vert \leq \delta}
\frac{\vert R_n\(\theta\)\vert}{1 + n\vert\theta-\theta_0\vert^2}
> \epsilon \Bigg \} < \epsilon.
\end{split}\end{align}
This stochastic equicontinuity condition  is equivalent to the
following stochastic equicontinuity condition, see e.g.
\cite{andrews1}:
 \bsnumber\label{pre_nm}
\text{ for any $\delta_n \rightarrow 0$ } \ \ \
\sup_{|\theta-\theta_0| \leq \delta_n} \frac{\vert
R_n\(\theta\)\vert}{1 + n\vert\theta-\theta_0\vert^2} =o_p(1).
 \end{split}\end{align}

 This is weaker than condition (v) of Theorem
7.1 in \cite{newey_mcfadden}, which requires \bsnumber\label{nm}
\sup_{\vert \theta-\theta_0\vert \leq \delta_n}
\frac{R_n\(\theta\)}{\sqrt{n}\vert \theta-\theta_0\vert + n\vert
\theta-\theta_0\vert^2} = o_p\(1\),
\end{split}\end{align}
since  \bs \frac{R_n\(\theta\)}{1 + n\vert \theta-\theta_0\vert^2}
= \frac{R_n\(\theta\)}{\sqrt{n}\vert \theta-\theta_0\vert + n\vert
\theta-\theta_0\vert^2}\[ \frac{\sqrt{n}\vert \theta-\theta_0\vert
+ n\vert \theta-\theta_0\vert^2}{1 + n\vert
\theta-\theta_0\vert^2}\],
\end{split}\end{align}
where the term in brackets is bounded by \bs 1 +
\frac{\sqrt{n}\vert \theta-\theta_0\vert}{1 + n\vert
\theta-\theta_0\vert^2} \leq 2.
\end{split}\end{align}
Hence the arguments of the proof, except for several important
differences, follow those of Theorem 7.2 in \cite{newey_mcfadden}.

At first note that condition \textbf{iv} of Proposition
\ref{prop1} is implied by the condition (where we let $g\(\theta\)
\equiv E g_n\(\theta\)$): \bsnumber\label{seq} \sup_{\theta \in
B_{\delta_n}(\theta_0) } \epsilon\(\theta\) =
o_p\(\frac{1}{\sqrt{n}}\), \text{ where } \epsilon\(\theta\) =
\frac{g_n\(\theta\) - g_n\(\theta_0\) - g\(\theta\)}{1 +
\sqrt{n}\vert\theta-\theta_0\vert} \text{ for any } \delta_n
\rightarrow 0.
\end{split}\end{align}
\noindent From (\ref{GMMid}) $$R_n\(\theta\)= R_{1n}(\theta) +
R_{2n}(\theta)+ R_{3n}(\theta),$$ where \bs R_{1n}
\(\theta\)\equiv &  n\biggl(  g_n\(\theta_0\)'W_n(\theta)
G\(\theta_0\)\(\theta-\theta_0\) + \frac12
\(\theta-\theta_0\)'G\(\theta_0\)'W(\theta)
G\(\theta_0\)\(\theta-\theta_0\) \\ &  \hspace{1in} - \frac12
g_n\(\theta\)' W_n(\theta) g_n\(\theta\) + \frac12
g_n\(\theta_0\)'
W_n(\theta) g_n\(\theta_0\) \biggr ), \\
 R_{2n}
\(\theta\)\equiv &   n\biggl(  \frac12 g_n\(\theta_0\)' \(
W_n(\theta_0) - W_n(\theta) \) g_n\(\theta_0\)  \biggr ),\\
 R_{3n}(\theta) \equiv &  n\biggl( g_n\(\theta_0\)'\(
 W(\theta_0) - W_n(\theta)) \)
G\(\theta_0\)\(\theta-\theta_0\) \\ &  \hspace{1in}  + \frac12
\(\theta-\theta_0\)'G\(\theta_0\)'\(  W(\theta_0)  - W(\theta) \)
G\(\theta_0\)\(\theta-\theta_0\)  \biggr ).
\end{split}\end{align}

Verification of (\ref{pre_nm}) for the terms $R_{2n}(\theta)$ and
$R_{3n}$ immediately follows from  $\sqrt{n} g_n(\theta_0) =
O_p(1)$ and the uniform consistency of $W_n(\theta)$ in $\theta$
as assumed in condition \textbf{i} of Proposition 1 and from the
continuity of $W(\theta)$ in $\theta$ by condition \textbf{i} of
Proposition 1, so that $W_n(\theta) - W(\theta) =o_p(1)$
uniformly in $\theta$ and $W(\theta) - W(\theta_0) =o(1)$ as
$|\theta - \theta_0| \rightarrow 0$.

 It remains to check condition (\ref{pre_nm}) for the
term $R_{1n}(\theta)$. Note that $$g_n\(\theta\) =
\(1+\sqrt{n}\vert
\theta-\theta_0\vert\)\epsilon\(\theta\)+g\(\theta\)+g_n\(\theta_0\).$$
Substitute this into $R_{1n}\(\theta\)$ and decompose \bs &
-\frac1n R_{1n}\(\theta\) =
\underbrace{\frac12\(1+\sqrt{n}\vert\theta-\theta_0\vert\)^2
\epsilon\(\theta\)'W_n(\theta)
\epsilon\(\theta\)}_{r_1\(\theta\)}+
\underbrace{g_n\(\theta_0\)'W_n(\theta) (g\(\theta\)-
G\(\theta_0\))\(\theta-\theta_0\)}_{r_2\(\theta\)}
\\
&+\underbrace{\(1+\sqrt{n}\vert \theta-\theta_0\vert\)
\epsilon\(\theta\)' W_n(\theta)
g_n\(\theta_0\)}_{r_3\(\theta\)}+\underbrace{\(1+\sqrt{n}\vert
\theta-\theta_0\vert\)
\epsilon\(\theta\)' W_n (\theta)g\(\theta\)}_{r_4\(\theta\)}\\
&+\underbrace{\frac12 g\(\theta\)'\(W_n (\theta)- W(\theta)\)
g\(\theta\)}_{r_5\(\theta\)}
 + \underbrace{\frac12 g\(\theta\)'W(\theta) g\(\theta\) -
\frac12 \(\theta-\theta_0\)'G\(\theta_0\)'W(\theta)
G\(\theta_0\)\(\theta-\theta_0\)}_{r_6\(\theta\)}.
\end{split}\end{align}

\noindent Using the inequalities, for $x \geq 0$:
\bsnumber\label{rel} \frac{\(1+\sqrt{n} x\)^2}{1 + nx^2} \leq 2,
\quad \frac{\sqrt{n} x}{1 + nx^2} \leq 1,\quad \frac{1+\sqrt{n}
x}{1 + nx^2} \leq 2,\quad \frac{n\(1+\sqrt{n}x\)}{1+nx^2} \leq 2
\frac{\sqrt{n}}{x},
\end{split}\end{align}
\noindent each of these terms can be dealt with separately, by
applying the conditions \textbf{i}- \textbf{iii} and (\ref{seq}):
 \bs (a) \ \ \ &  \sup_{\theta \in B_{\delta_n}(\theta_0)}
\frac{n |r_1\(\theta\)|}{1 + n\vert\theta-\theta_0\vert^2} {\leq}
\sup_{\theta \in B_{\delta_n}(\theta_0)} n \epsilon\(\theta\)'
W_n (\theta)
\epsilon\(\theta\) {=}o_p\(1\), \\
(b) \ \ \ &  \sup_{\theta \in B_{\delta_n}(\theta_0)} \frac{n
|r_2\(\theta\)|}{1 + n\vert\theta-\theta_0\vert^2} {=}
\sup_{\theta \in B_{\delta_n}(\theta_0)}\frac{
o\(\sqrt{n}\vert\theta-\theta_0\vert\)'}{1 +
n\vert\theta-\theta_0\vert^2} W_n (\theta) \sqrt{n}
g_n\(\theta_0\) = o_p\(1\), \\
(c) \ \ \ & \sup_{\theta \in B_{\delta_n}(\theta_0)}  \frac{n
|r_3\(\theta\)|}{1 + n\vert\theta-\theta_0\vert^2}\leq
\sup_{\theta \in B_{\delta_n}(\theta_0)}
2 n| \epsilon\(\theta\)' W_n (\theta)g_n\(\theta_0\) |=o_p\(1\),\\
(d) \ \ \ & \sup_{\theta \in B_{\delta_n}(\theta_0)} \frac{n
|r_4\(\theta\)|}{1 + n\vert\theta-\theta_0\vert^2} \leq
\sup_{\theta \in B_{\delta_n}(\theta_0)} 2 \sqrt{n} \left |
\epsilon\(\theta\)'
W_n (\theta) \frac {g\(\theta\)} {|\theta-\theta_0|} \right | = o_p\(1\),\\
(e) \ \ \ & \sup_{\theta \in B_{\delta_n}(\theta_0)}\frac{ n
|r_5\(\theta\)|}{1 + n\vert\theta-\theta_0\vert^2} \leq
\sup_{\theta \in B_{\delta_n}(\theta_0)} \frac{\vert
g\(\theta\)\vert^2}{\vert \theta-\theta_0\vert^2} \vert W_n
(\theta)- W(\theta)\vert = o_p\(1\), \\
(f) \ \ \ & \sup_{\theta \in B_{\delta_n}(\theta_0)}\frac{ n
|r_6\(\theta\)|}{1 + n\vert\theta-\theta_0\vert^2} \leq
\sup_{\theta \in B_{\delta_n}(\theta_0)} \frac{o\(
|\theta-\theta_0|^2 | W(\theta)|  \)}{| \theta-\theta_0|^2} =
o_p\(1\),
 \end{split}\end{align}
where (a) follows from (\ref{rel}), (\ref{seq}), and condition
\textbf{i}, which states that $W_n (\theta) = W(\theta) +o_p(1)$
and $W(\theta)>0$ is finite uniformly in $\theta$; in (b) the
first equality follows by Taylor expansion
$g(\theta)=G\(\theta_0\)\(\theta-\theta_0\)+o\(\vert\theta-\theta_0\vert\)$
and the second conclusion follows from (\ref{rel}) and condition
\textbf{iii}; (c) follows from (\ref{rel}), (\ref{seq}),
condition \textbf{i} and \textbf{iii}; (d) follows by (\ref{rel})
and then replacing, by condition \textbf{ii}, $g\(\theta\)$ with
$G\(\theta_0\)\(\theta-\theta_0\)+o\(\vert\theta-\theta_0\vert\)$,
followed by applying (\ref{seq}) and condition \textbf{i}; (e)
follows from replacing by condition \textbf{ii} $g\(\theta\)$ with
$G\(\theta_0\)\(\theta-\theta_0\)+o\(\vert\theta-\theta_0\vert\)$,
followed by applying condition \textbf{i}; and (f) follows from
replacing $g\(\theta\)$ with
$G\(\theta_0\)\(\theta-\theta_0\)+o\(\vert\theta-\theta_0\vert\)$,
followed by applying condition \textbf{i}.

Verification of (\ref{nm}) for the term $R_{1n}(\theta)$ now
follows by putting these terms together.\hfill $\blacksquare$

\subsection{Proof of Proposition 2}
Verification of Assumption 3 is standard given the stated
conditions and is subsumed as a step in the consistency proofs of
extremum estimators based on GEL in \cite{kitamura_stutzer} for
cases when $s$ is finite and \cite{kitamura:annals} for cases
when $s$ takes on infinite values. We shall not repeat it here.
Next, we will verify Assumption 4. Define \bs \widehat
\gamma\(\theta\) = \arg\inf_{\gamma\in\Bbb{R}^p} \bar L_n\(\theta,
\gamma\).
\end{split}\end{align}
It will suffice to show that uniformly in $\theta_n \in
B_{\delta_n}(\theta_0)$ for any $\delta_n \rightarrow 0$, we have
the GMM set-up:
 \bsnumber\label{gel_bs} L_n\(\theta_n\)& = \bar L_n\(\theta_n,
\widehat \gamma\(\theta_n\)\) \\ & = -
\frac{1}{2}\(\frac{1}{\sqrt{n}} \sum_{i=1}^n m_i\(\theta_n\)\)'
\(V\(\theta_0\)+o_p\(1\)\)^{-1}\(\frac{1}{\sqrt{n}} \sum_{i=1}^n
m_i\(\theta_n\)\),
\end{split}\end{align}
where
$$
V(\theta_0) =  E  m_i\(\theta_0\) m_i\(\theta_0\)'.
$$
The Assumptions 4.\textbf{i-iii} follow immediately from the
conditions of Proposition 2,  and Assumption 4.\textbf{iv} is
verified exactly as in the proof of Proposition 1, given the
reduction to the GMM case. Indeed, defining $g_n(\theta) =
\frac{1}{n} \sum_{i=1}^n m_i(\theta)$, the Donsker property
assumed in condition \textbf{iv} implies that for any
$\epsilon>0$, there is $\delta>0$ such that
$$
\limsup_{n \rightarrow \infty } P^* \Bigg \{ \sup_{\theta \in
B_{\delta} (\theta_0) } {\sqrt{n} | g_n(\theta) - g_n(\theta_0) -
( E g_n(\theta) - E g_n(\theta_0)) |} > \epsilon\Bigg \} <
\epsilon,
$$
which implies
$$
\limsup_{n \rightarrow \infty } P^* \Bigg \{ \sup_{\theta \in
B_{\delta} (\theta_0) } \frac{\sqrt{n} | g_n(\theta) -
g_n(\theta_0) - ( E g_n(\theta) - E g_n(\theta_0)) |}{1 +
\sqrt{n} |\theta- \theta_0| }
> \epsilon\Bigg \} < \epsilon,
$$
which is  condition \textbf{iv} in Proposition 1. The rest of the
arguments follow that in the proof of Proposition 1.

It only remains to show the requisite expansion (\ref{gel_bs}).
We first show that
$$
\widehat \gamma(\theta_n) \inp 0.
$$
For that purpose we use the convexity lemma, which was obtained
by C. Geyer, and can be found in \cite{knight}.

\noindent \textbf{Convexity Lemma.} Suppose $Q_n$ is a  sequence
of lower-semi-continuous convex $\bar{\Bbb{R}}$-valued random
functions, defined on $\Bbb{R}^d$, and let $\mathcal{D}$ be a
countable dense subset of $\Bbb{R}^d$. If $Q_n$ weakly converges
to $Q_{\infty}$ in $\bar{\Bbb{R}}$ marginally (in
finite-dimensional sense) on $\mathcal{D}$ where $Q_{\infty}$ is
lower-semi-continuous convex and finite on an open non-empty set
a.s., then
$$ \underset{z\in \Bbb{R}^d}{\arg \inf }\ {Q_n(z)} \ind
\underset{z\in \Bbb{R}^d}{\arg \inf }\ {Q_{\infty}(z)},$$
provided the latter is uniquely defined a.s. in $\Bbb{R}^d $.

Next, we show that $ \widehat \gamma(\theta_n) \inp 0$.  Define
$F = \{ \gamma: E s [ m_i(\theta_0)'\gamma] < \infty\}$ and $F^c
= \{ \gamma: E s [ m_i(\theta_0)'\gamma]= \infty\}$. By convexity
and lower-semicontinuity of $s$, $F$ is
 convex, open, and its boundary is nowhere dense in
 $\Bbb{R}^{p}.$ Thus for
 $\gamma \in F$, $E s [ m_i(\theta)'\gamma]
 < \infty$ for all $\theta\in B_{\delta}(\theta_0)$ and some
 $\delta>0$,  which follows by continuity of $\theta
\mapsto E s [ m_i(\theta)'\gamma]$ over $B_{\delta}(\theta_0)$
implied by the condition \textbf{ii} and \textbf{iii}.

Thus, for a given  $\gamma \in F$ and any $\theta_n \inp \theta_0$
 $$\frac{1}{n} \sum_{i=1}^n s [ m_i(\theta_n)'\gamma] \inp E s [ m_i(\theta_0)'\gamma] < \infty.$$
 This follows  from the uniform law of large numbers implied by

\textbf{1.}
 $
\{ s [ m_i(\theta)'\gamma ], \theta\in B_{\delta}(\theta_0) \},
\text{ where }  \delta$ is sufficiently small, being a Donsker
class wp $\rightarrow 1$, and

\textbf{2.} $E m_{ij}(\theta)= \int x \cdot d P[ m_{ij}(\theta) \leq x]
$ being continuously differentiable in $\theta$ [in a fixed open ball containing $\theta_0$] by assumption.

\noindent The above function set is Donsker by
\begin{itemize}
\item[(a)] $m_i(\theta)'\gamma \in M$ for some compact $M$
and a given $\gamma \in F$, by condition \textbf{iii},
\item[(b)] $\{ m_i(\theta), \theta\in B_{\delta}(\theta_0) \}$ being
 Donsker class by condition
\textbf{iv},
\item[(c)] $s$ being a uniform Lipschitz function over
$\mathcal{V} \cap M$\footnote{Recall that $\mathcal{V}$ is
defined as the open convex set on which $s$  is finite. }, by
assumption on $s$,
\item[(d)]  $m_i(\theta)'\gamma \in
\mathcal{V}$ for all $\theta\in B_{\delta}(\theta_0)$, some
$\delta>0$, and a given $\gamma \in F$, by construction of $F$,
\item[(e)] Theorem 2.10.6 in \cite{vaart} that says a
uniform Lipschitz transform of a Donsker class is Donsker class
itself.
 \end{itemize}

Now take $\gamma$ in $F^c\setminus \partial F$, where $\partial F$
denotes the boundary of $F$. Then $\wpa 1$
$$
\frac{1}{n} \sum_{i=1}^n s [ m_i(\theta_n)'\gamma]= \infty \inp E
s [ m_i(\theta_0)'\gamma] = \infty.
$$
 Now take all the rational numbers $\gamma \in \Bbb{R}^p \setminus \partial F$
  as the set $\mathcal{D}$ appearing in the
statement of the Convexity Lemma and conclude that
$$
\widehat \gamma(\theta_n) \inp 0 = \text{arg inf}_{\gamma} E s [
m_i(\theta_0)'\gamma].
$$

Given this result, we can expand the first order condition for
$\widehat \gamma(\theta_n)$ in order to obtain the expression for
its form. Note first
 \bsnumber\label{expandfoc} 0 & = \sum_{i=1}^n \nabla
s\(\gamma\(\theta_n\)'m_i\(\theta_n\)\) m_i\(\theta_n\) \\
& = \sum_{i=1}^n m_i\(\theta_n\) + \gamma\(\theta_n\) n V_n,
 \end{split}\end{align}
where
$$
V_n = \frac{1}{n}\sum_{i=1}^n \nabla^2 s\(\bar
\gamma\(\theta_n\)' m_i\(\theta_n\)\) m_i\(\theta_n\)
m_i\(\theta_n\)',
$$
for some $\bar\gamma\(\theta_n\)$ between $0$ and
$\gamma\(\theta_n\)$, which is different from row to row of the
matrix $V_n$.

Then
$$
V_n \inp V(\theta_0) = E  m_i\(\theta_0\) m_i\(\theta_0\)'.
$$
 This follows  from the uniform law of large numbers  implied by

\textbf{1.}
 $
\{ \nabla^2 s\(\gamma' m_i\(\theta^*\)\) m_i\(\theta\)
m_i\(\theta\)', (\theta^*, \gamma, \theta) \in
B_{\delta_1}(\theta_0)\times B_{\delta_2}(0)\times B_{\delta_3}
(\theta_0)\}, \text{ where } \delta_{j} > 0$ are sufficiently
small, being a Donsker class wp $\rightarrow 1$,

\textbf{2.} $E m_i(\theta) m_i(\theta)'$ being continuous function in $\theta$ by
assumption,

\textbf{3.} $E\nabla^2 s( \gamma' m_i\(\theta^*\)) m_i\(\theta\)
m_i\(\theta\)' =  E {\sss \underset{1}{\underbrace{\nabla^2
s(0)}}} m_i\(\theta\) m_i\(\theta\)' + o(1)$ uniformly in
$(\theta, \theta^*) \in B_{\delta}(\theta_0)\times
B_{\delta}(\theta_0)$  for sufficiently small $\delta >0$, for
any $\gamma \rightarrow 0$, by assumptions on $s$ and condition
\textbf{iii}.

\noindent The claim \textbf{1} is verified by applying exactly the
same logic as in the previously stated steps (a)-(e). For the
sake of brevity, this will not be repeated.

Therefore, $\wpa 1$ \bsnumber\label{gel_bs2}
 \gamma\(\theta_n\) = - \( V_n\)^{-1} \frac1n \sum_{i=1}^n
m_i\(\theta_n\)\equiv - (V(\theta_0)^{-1} +o_p(1)) \frac1n
\sum_{i=1}^n m_i\(\theta_n\).
\end{split}\end{align}
Consider the second order expansion,
 \bsnumber\label{gel_bs3}
\bar L_n\(\theta_n, \gamma\(\theta_n\)\) =&
\sqrt{n}\gamma\(\theta_n\)' \frac{1}{\sqrt{n}} \sum_{i=1}^n
m_i\(\theta_n\)  +\frac12 \sqrt{n}\gamma\(\theta_n\)' \tilde V_n
\sqrt{n}\gamma\(\theta_n\),
\end{split}\end{align}
where
$$
\tilde V_n = \frac{1}{n}\sum_{i=1}^n \nabla^2 s\(\tilde
\gamma\(\theta_n\)' m_i\(\theta_n\)\) m_i\(\theta_n\)
m_i\(\theta_n\)',
$$
for some $\tilde\gamma\(\theta_n\)$ between $0$ and
$\gamma\(\theta_n\)$, which is different from row to row of the
matrix $\tilde V_n$.  By a preceding argument,
$$
\tilde V_n \inp V(\theta_0).
$$
Inserting  (\ref{gel_bs2}) and $\tilde V_n = V(\theta_0)+o_p(1)$
into (\ref{gel_bs3}), we obtain the required expansion
(\ref{gel_bs}). $\blacksquare$

\subsection{Proof of Proposition 3}  Assumption \ref{consistency assumption}
is assumed. We need to verify Assumption \ref{second order
expansion}.

Define the identity
 \bsnumber\label{Mid}
L_n(\theta) - L_n(\theta_0) & \equiv
\underset{\Delta_n(\theta_0)'}{\underbrace {\sum_{i=1}^n \dot
m_i(\theta_0)'}} (\theta - \theta_0) \\
& + \frac12 (\theta - \theta_0)' n
\underset{-J(\theta_0)}{\underbrace {\nabla_{\theta \theta'}
E m_i(\theta_0)}}(\theta - \theta_0) \\
&+ R_n(\theta).
\end{split}\end{align}
Assumption 4.\textbf{i-iii} then follows immediately from
conditions \textbf{i} and \textbf{ii}. Assumption 4.\textbf{iv}
is verified as follows.

The remainder term $R_n\(\theta\)$ is given the following
decomposition: \bs R_n\(\theta\) =& \underbrace{\sum_{i=1}^n \Big
\{ m_i (\theta) - m_i (\theta_0) -  E m_i (\theta) + E m_i
(\theta_0)  - \dot m_i
(\theta_0)' (\theta - \theta_0) \Big \}}_{R_{1n}\(\theta\)}\\
&\hspace{1in} + \underbrace{n\(E m_i (\theta) - E m_i
(\theta_0)\)  + \frac{1}{2}(\theta - \theta_0)' n J(\theta_0)
(\theta - \theta_0)}_{R_{2n}\(\theta\)}.
\end{split}\end{align}
It suffices to verify Assumption 4.\textbf{iv} separately for
$R_{1n}(\theta)$ and $R_{2n}(\theta)$. Since
 \bs
R_{2n}(\theta) & = - \frac{1}{2} n (\theta - \theta_0)' \[
J(\theta^*) - J(\theta_0) \] ( \theta- \theta_0),
\end{split}\end{align}
for some $\theta^*$ on the line connecting $\theta$ and
$\theta_0$,  verification of Assumption 4 for $R_{2n}(\theta)$
follows immediately from continuity of  $J(\theta)$ in $\theta$
over a ball at $\theta_0$.

\noindent To show Assumption \ref{second order
expansion}.\textbf{iv}-(b) for $R_{1n}(\theta)$, we note that for
any given $M >0$
 \bsnumber\label{p.mp.1} \limsup_n  P^* &
\left \{ \sup_{  |\theta - \theta_0 | \leq
M/\sqrt{n}}\ \ \ \  | R_{1n}\(\theta\) | > \epsilon \right \} \\
& \leq \limsup_n  P \left \{ \sup_{  |\theta - \theta_0 | \leq
M/\sqrt{n}}\ \ \ \   |\theta - \theta_0| \frac{ |
R_{1n}\(\theta\)| } { | \theta - \theta_0 | }
> \epsilon  \right \} \\
 & \leq \limsup_n  P \left \{ \sup_{  |\theta - \theta_0 | \leq
M/\sqrt{n}}\ \ \ \   \frac {M} {\sqrt{n}} \frac{ |
R_{1n}\(\theta\)| }  { | \theta - \theta_0 | }
> \epsilon  \right \}
= 0,
\end{split}\end{align}
where the last conclusion follows from two observations.

\noindent First, note that
$$Z_n (\theta)  \equiv \frac {R_{1n}\(\theta\)}  { \sqrt{n}
| \theta - \theta_0 | } = \frac{1}{\sqrt{n}} \sum_{i=1}^n
\(\frac{m_i(\theta)- m_i(\theta_0) - (E m_i (\theta) - E m_i
(\theta_0)) - \dot m_i(\theta_0)' (\theta -
 \theta_0)}{| \theta - \theta_0 |}    \)
$$
is Donkser by assumption, that is it converges in $\ell^{\infty} (
B_{\delta}(\theta_0) )$ to a tight Gaussian process $Z$.

The process has uniformly continuous paths with respect to the
semimetric $\rho$ given by
$$
\rho^2 (\theta_1, \theta_2) = E ( Z(\theta_1) - Z(\theta_2) )^2,
$$
so that $\rho(\theta, \theta) \rightarrow 0$ if $\theta
\rightarrow \theta_0$.  Thus almost all sample paths of $Z$ are
continuous at $\theta_0$.

\noindent Second, since by assumption
$$
 E [ \overline{m}_n(\theta) - \overline{m}_n (\theta_0) -
\overline{\dot m_n}(\theta_0)'(\theta - \theta_0)]^2 = o( |
\theta- \theta_0|^2),
$$
we have for any  $\theta_n \rightarrow \theta_0$
$$
E^* \[  \frac {| R_{1n}\(\theta_n\)| }  { \sqrt{n} | \theta_n -
\theta_0 |} \]^2  =  \frac{ o\(| \theta_n - \theta|^2\) } {
|\theta_n - \theta_0|^2 } \rightarrow 0,
$$
therefore
$$
Z(\theta_0) = 0.
$$
Therefore for any $\theta' \inp \theta_0$, we have by the extended
continuous mapping theorem \bsnumber\label{p.mp.2} Z_n (\theta')
\ind Z(\theta_0) = 0, \text{ that is } Z_n (\theta') \inp 0.
\end{split}\end{align}
This shows (\ref{p.mp.1}).

\noindent To prove Assumption \ref{second order
expansion}.\textbf{iv}-(a) for $R_{1n}(\theta)$, we need to show
that for some $\delta > 0$ and constant $M$
 \bsnumber\label{p.mp.3}
\ \ \limsup_n \ &P^* \left \{ \sup_{  M/\sqrt{n}\leq |\theta -
 \theta_0 | \leq \delta } \frac{ | R_{1n}\(\theta\) |} { n | \theta -
 \theta_0 |^2}> \epsilon \right \} < \epsilon.
\end{split}\end{align}
Using that $ M/\sqrt{n}\leq |\theta - \theta_0 |$, bound the
left-hand-side by
 \bsnumber\label{p.mp.4}
\limsup_n \ & P^* \left \{ \sup_{  M/\sqrt{n}\leq |\theta -
 \theta_0 | \leq \delta } \frac{ | R_{1n}\(\theta\) |} { \sqrt{n} | \theta -
 \theta_0 |} \cdot \frac { 1} {\sqrt{n} |\theta- \theta_0| }> \epsilon \right
 \} \\
& \leq \limsup_n \  P^* \left \{ \sup_{  M/\sqrt{n}\leq |\theta -
 \theta_0 | \leq \delta } \frac{ | R_{1n}\(\theta\) |} { \sqrt{n} | \theta -
 \theta_0 |} \cdot \frac { 1 } { M }> \epsilon \right
 \} \\
& \leq \limsup_n \  P^* \left \{ \sup_{  M/\sqrt{n}\leq |\theta -
 \theta_0 | \leq \delta } | Z_n (\theta)| \cdot \frac { 1 } { M }> \epsilon \right
 \} \\
& < \epsilon,
\end{split}\end{align}
where for any given $\epsilon >0$ in order to make the last
inequality true, we can make either $\delta$ sufficiently small by
the property (\ref{p.mp.2}) of $Z_n$ or make $M$ sufficiently
large by the property $Z_n = O_{p^*}(1)$. \hfill
 $\blacksquare$

\section{Appendix on Computation}\label{implementation}

\subsection{A computational lemma} In this section we record some
formal results on MCMC computation of the quasi-posterior
quantities.
\begin{lemma}\label{mh_lemma}  Suppose the chain $(\theta^{(j)},
j \leq B)$ is produced by the Metropolis Hastings(MH) algorithm
with $q$ such that $q(\theta |\theta')
>0$ for each $(\theta,\theta')$. Suppose also that $P\{ \rho(
\theta^{(j)},\xi) =1\} >0$ for all  $j > t_0$. Then
\begin{itemize}
 \item[1.]  $p_n(\cdot)$ is the stationary density of the
 chain,
 \item[2.] the chain is ergodic with the limit marginal
 distribution given by $p_n(\cdot)$:
{\bs \lim_{B \rightarrow \infty} \sup_{A} \Big | P(\theta^{(B)}
\in A | \theta_0) - \int_A p_n(\theta) d \theta \Big | =0,
 \end{split}\end{align}}
 where the supremum is taken over the Borel sets,
 \item[3.] For any $p_n$- integrable function $g$:
 {\bs
\frac{1}{B} \sum_{j=1}^B g(\theta^{(j)}) \inp \int_{\Theta}
g(\theta)  p_n(\theta) d\theta.
 \end{split}\end{align}}
\end{itemize}
\end{lemma}
\noindent \textbf{Proof.} The result is immediate from Theorem
6.2.5 in \cite{montecarlo}. $\blacksquare$

An immediate consequence of this lemma is the following result.

\begin{lemma}\label{chain converge} Suppose Assumptions 1 and 2 hold.
Suppose the chain $(\theta^{(j)}, j \leq B)$ satisfies the
conditions of Lemma \ref{mh_lemma}, then for any convex and
$p_n$-integrable loss function $\rho_n\(\cdot\)$ \bs
\arg\inf_{\theta \in \Theta} \bigg [\frac{1}{B} \sum_{j=1}^B
\rho_n (\theta^{(j)} - \theta)\bigg ] \inp \widehat \theta =
\arg\inf_{\theta \in \Theta} \bigg [\int_{\Theta} \rho_n ( \tilde
\theta - \theta) p_n(\tilde \theta) d \tilde \theta\bigg],
\end{split}\end{align}
provided that $\widehat \theta$ is uniquely defined.
\end{lemma}

\noindent \textbf{Proof.} By Lemma \ref{mh_lemma} we have the
pointwise convergence of the objective function: for any $\theta$
$$
\frac{1}{B} \sum_{j=1}^B \rho_n (\theta^{(j)} - \theta) \inp
\int_{\Theta} \rho_n (\tilde \theta - \theta) p_n(\tilde \theta) d
\tilde \theta,
$$
which implies the result by the Convexity Lemma, since $\theta
\mapsto \int_{\Theta} \rho_n (\tilde \theta - \theta) p_n(\tilde
\theta) d \tilde \theta$ is convex by convexity of $\rho_n$.
$\blacksquare$

%The Metropolis-Hastings algorithm has a very intuitive structure.
%It generates a time-homogeneous Markov chain with the transition
%kernel given by:
%$$
% K(x|y) = \rho(x, y) q (y|x) +   (1 - r(x)) \delta_x(y),
%$$
%where $r(x)  = \int \rho(x, y) q(y|x) d y$ and $\delta_x$ denotes
%the Dirac mass at $x$.  The transition kernel $K$ satisfies the
%detailed balance condition (DBC) relative to $p_n$:
%$$
%K(y,x) p_n(y) = K(x, y) p_n (x),
%$$
%for every $(x,y)$. This condition states that the move from $y$ to
%$x$ is as likely as the move form $x$ to $y$ when the initial
%point is drawn from $p_n$. The DBC is an elegant sufficient
%condition for $p_n$ to be a stationary distribution of the chain:
%$$
%Pr ( x \in A) = \int_A \int_{\Theta} K(x,y) p_n(x) d x dy
% = \int_{A} \int_{\Theta} K(y,x) p_n(y) d x dy = \int_{A} p_n(y)
% dy.
%$$
%The conditions on $q$ require the chain to explore the support of
%$p_n$ and are sufficient for $p_n$-irreducibility.  A sufficient
%condition for the MH-chain to be aperiodic is that the algorithm
%allows the events such as $\{\theta^{(j+1)} = \theta^{(j)}\}$
%with non-zero probability.

\subsection{Quasi-Bayes Estimation and Simulated
Annealing}\label{annealing}

The relation between drawing from the shape of a likelihood
surface and optimizing to find the mode of the likelihood function
is well known. It is well established that, e.g.
\cite{montecarlo},
 \bsnumber\label{limit is optimum}
\lim_{\lambda\rightarrow\infty} \frac{\int_\Theta \theta
e^{\lambda L_n\(\theta\)} \pi\(\theta\) d\theta}{ \int_\Theta
e^{\lambda L_n\(\theta\)} \pi\(\theta\) d\theta} =
\underset{\theta\in\Theta}{\text{argmax}} L_n\(\theta\)
\end{split}\end{align}
Essentially, as $\lambda \rightarrow \infty$, the sequence of
probability measures \bsnumber\label{bayesdirac} \frac{e^{\lambda
L_n\(\theta\)}\pi\(\theta\)}{\int_\Theta e^{\lambda L_n\(\theta\)}
\pi\(\theta\) d\theta}
\end{split}\end{align}
converges to the generalized Dirac probability measure
concentrated at $\underset{\theta\in\Theta}{\text{argmax}} \
L_n\(\theta\)$.

The difficulty of nonlinear optimization has been an important
issue in econometrics (\cite{bhhh}, \cite{bayesmth}). The
simulated annealing algorithm (see e.g. \cite{numerical},
\cite{simulatedannealing}) is usually considered a generic
optimization method. It is an implementation of the simulation
based optimization (\ref{limit is optimum}) with a uniform prior
$\pi\(\theta\) \equiv c$ on the parameter space $\Theta$. At each
temperature level $1/\lambda$, the simulated annealing routine
uses a large number of Metropolis-Hastings steps to draw from the
quasi distribution (\ref{bayesdirac}). The temperature parameter
is then decreased slowly while the Metropolis steps are repeated,
until convergence criteria for the optimum are achieved.

Interestingly, the simulated annealing algorithm has been widely
used in optimization of non-likelihood-based semiparametric
objective functions. In principle, if the temperature parameter is
decreased at an arbitrarily slow rate (that depends on the
criterion function), simulated annealing can find the global
optimum of non-smooth objective functions that may have many local
extrema. Controlling the temperature reduction parameter is a very
delicate matter and is certainly crucial to the performance of the
algorithm with highly nonsmooth objective functions. On the other
hand, as Theorems \ref{theorem2} and \ref{theorem3} apply equally
to (\ref{bayesdirac}), the results of this paper show that we may
fix the temperature parameter $1/\lambda$ at a positive constant
and then compute the quasi-posterior medians or means for
(\ref{bayesdirac}) using Metropolis steps.  These estimates can be
used in place of the exact maximum. They are consistent and
asymptotically normal, and possess the same limiting distribution
as the exact maximum. The interpretation of the simulated
annealing algorithm as an implementation of (\ref{bayesdirac})
also suggests that for some problems with special structures,
other MCMC methods, such as the Gibbs sampler, may be used to
replace the Metropolis-Hasting step in the simulated annealing
algorithm.

\subsection{Details of Computation in Monte-Carlo Examples}
The parameter space is taken to be $\Theta = [ \theta_0 \pm 10]$.
The transition kernel is a Normal density, and  flat prior is
truncated to $\Theta$. Each parameter is updated via a
Gibbs-Metropolis procedure, which modifies slightly the basic
Metropolis-Hastings algorithm: for $k = 1,...,d$, a draw of
$\xi_k$ from the univariate normal density $q(|\xi_k -
\theta^{(j)}_k|, \phi)$ is made, then the candidate value $\xi$
consisting of $\xi_k$ and $\theta^{(j)}_{-k}$ replaces
$\theta^{(j)}$ with probability $\rho(\theta^{(j)}, \xi)$
specified in the text. Variance parameter $\phi$ is adjusted
every 100 draws (in the second simulation example and empirical
example) or 200 draws (in the first simulation example) so that
the rejection probability is roughly $50\%$.

 The first $N\times d$ draws  (the
burn-in stage) are discarded, and the remaining $N\times d$ draws
are used in computation of estimates and intervals. The starting
value is the OLS estimate in all examples. We use $N=5,000$ in
the second simulation example and empirical example and
$N=10,000$ in the second simulation example. To give an idea of
computational expense, computing one set of estimates takes 20-40
seconds depending on the example. All of the codes that we used to
produce figures, simulation, and empirical results are available
from the authors.

\section*{Notation and Terms}
\begin{tabular}{rl}
  $\inp$  & convergence in (outer) probability $P^*$\\
  $\ind$ & convergence in distribution  under $P^*$ \\
  wp $\rightarrow 1$  & with inner probability $P_*$ converging to one \\
  $\sim$ & asymptotic equivalence denoted $A \sim B$ means $\lim
  A B^{-1} =I$ \\
  $B_{\delta}(x)$ & ball centered at $x$ of radius $\delta>0$ \\
  $I$ & identity matrix \\
  $A>0$ &  $A$ is positive definite when $A$ is matrix \\
  $\mathcal{N}(0, a)$ & normal random vector with mean $0$ and
  variance matrix $a$ \\
  $\mathcal{F}$ Donsker class & here this means that empirical
  process $f \mapsto \frac{1}{\sqrt{n}}
  \sum_{i=1}^n (f(W_i) - E f (W_i))$ is \\
  &asymptotically Gaussian in $\ell^{\infty}(\mathcal{F})$, see
  \cite{vandervaart}\\
 % & \\
  $\ell^{\infty}(\mathcal{F})$ & metric space of bounded over $\mathcal{F}$
  functions, see \cite{vandervaart} \\
  \text{mineig}(A) & minimum eigenvalue of matrix $A$
    \end{tabular}

\newpage

 \nocite{owen:1989} \nocite{owen:1990}
\nocite{owen:1991} \nocite{owen_book}

\newpage

\section*{Tables and Figures}

\begin{table}[ht]\caption{Monte Carlo  Comparison of LTE's with
Censored Quantile Regression Estimates Obtained using Iterated
Linear Programming (Based on 100 repetitions).}\label{table1}
\begin{center}
\begin{tabular}{l|c|c|c|c|c}
\hline\hline
 Estimator \hspace{.3in}  & RMSE & MAD & Mean Bias & Median Bias  & Median Abs. Dev.\\
 \hline
  \textbf{n=400 }& & &  & & \\

 Q-posterior-mean                       &  0.473 & 0.378 & 0.138 & 0.134 & 0.340 \\
 Q-posterior-median                     &  0.465 & 0.372  & 0.131 & 0.137 & 0.344\\
Iterated LP(10) & 0.518 & 0.284 & 0.040 & 0.016 &  0.171 \\
 & 3.798 & 0.827 & -0.568 & -0.035 &  0.240 \\
  \textbf{n=1600 } & & &  &  &\\
 Q-posterior-mean                       & 0.155  & 0.121 & -0.018 & 0.009 & 0.089\\
 Q-posterior-median                     & 0.155  & 0.121  & -0.020 & 0.002 & 0.092\\
Iterated LP(7) & 0.134 & 0.106 & 0.040 & 0.067 & 0.085\\
 & 3.547 & 0.511 & 0.023 & -0.384 & 0.087 \\\hline\hline
\end{tabular}
\end{center}
\end{table}

\begin{table}
\caption{Monte Carlo Comparison of the LTE's with Standard
Estimation for a Linear Quantile Regression Model (Based on 500
repetitions)}\label{table2}
\begin{center}
\begin{tabular}{l|c|c|c|c|c }
\hline\hline
 Estimator & RMSE & MAD & Mean Bias & Median Bias  &  Median AD   \\
 \hline
  \textbf{n=200 }& & &  & & \\
 Q-posterior-mean                       & .0747 & .0587 & .0174 &
.0204 & .0478 \\
 Q-posterior-median                     & .0779 & .0608  & .0192 & .0136 & .0519\\
 Standard Quantile Regression & .0787 & .0628 &.0067 & .0092 & .0510\\
  \textbf{n=800 }& & &  &  &\\

 Q-posterior-mean                       & .0425 & .0323 & -.0018 & -.0003 & .0280\\
 Q-posterior-median                     & .0445 & .0339  & -.0023 & .0001 & .0295\\
 Standard Quantile Regression & .0498 & .0398 & .0007 & .0025 & .0356\\ \hline \hline
\end{tabular}
\end{center}
\end{table}

\begin{table}\caption{Monte Carlo  Comparison of the LT Inference with Standard
Inference for a Linear Quantile Regression Model (Based on 500
repetitions)}\label{table3}\begin{center}
\begin{tabular}{l|c|c }
 \hline\hline
 Inference Method  & \textbf{coverage} &\textbf{ length}      \\
 \hline
  \textbf{n=200 }& &   \\
 Quasi-posterior confidence interval, equal tailed & .943 & .377  \\
 Quasi-posterior confidence interval, symmetric (around mean)  & .941 & .375   \\
  Quantile Regression: Hall-Sheather Interval& .659 & .177 \\
  \textbf{n=800 }& &   \\
 Quasi-posterior confidence interval, equal tailed & .920 & .159  \\
 Quasi-posterior confidence interval, symmetric (around mean)  & .917 & .158   \\
 Quantile Regression: Hall-Sheather Interval& .602 & .082 \\
 \hline\hline
\end{tabular}
\end{center}\end{table}

\vspace{.05in}

\begin{figure}[htbp!] \label{Figure 2}
 \begin{center}
\noindent
 \includegraphics[width=5in,height=3in]{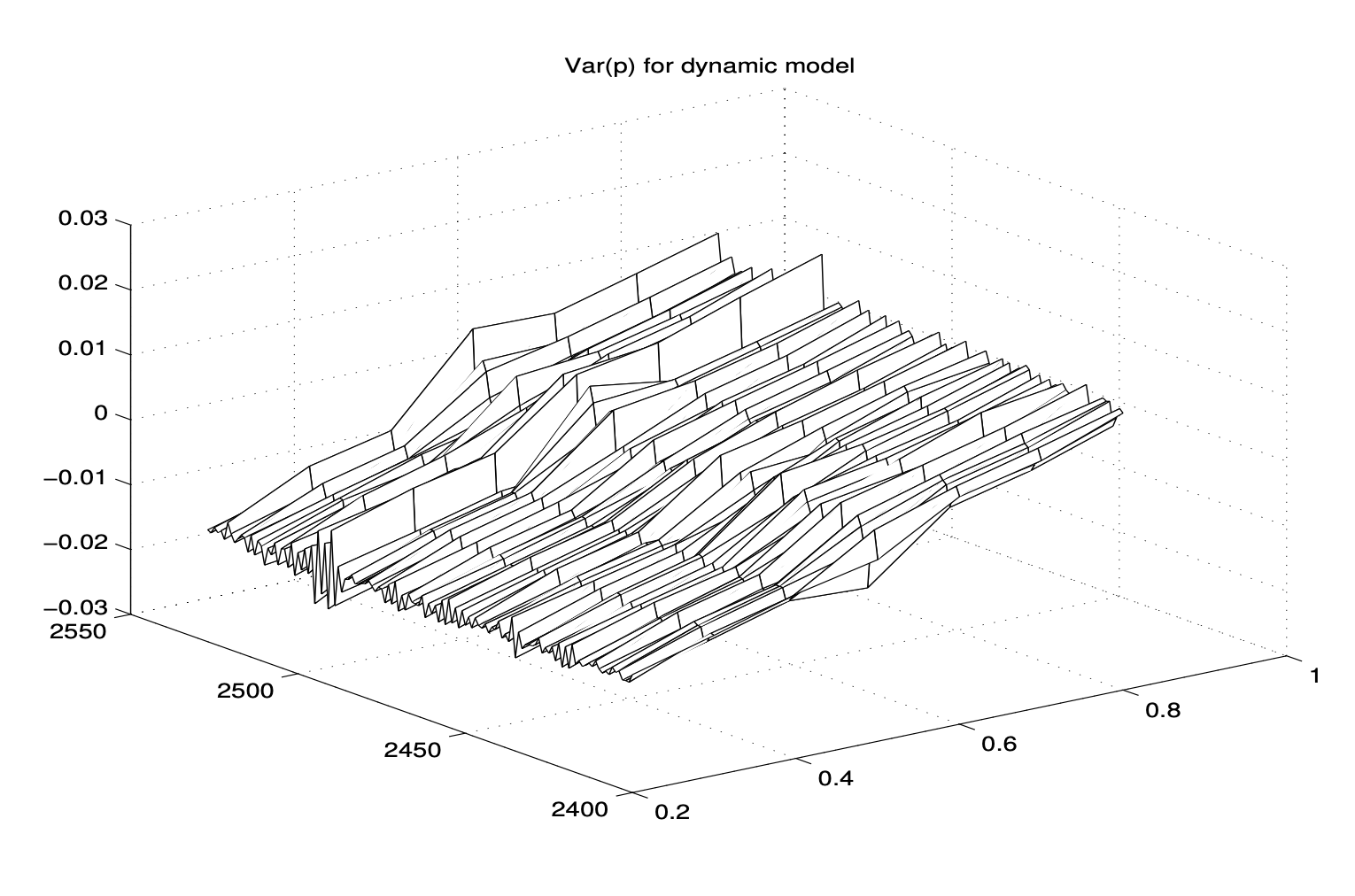}\caption{Recursive VaR Surface in time-probability space}
 \end{center}
\end{figure}

\begin{figure}[htbp!] \label{Figure 10}
 \begin{center}
\noindent
 \includegraphics[width=5in,height=3in]{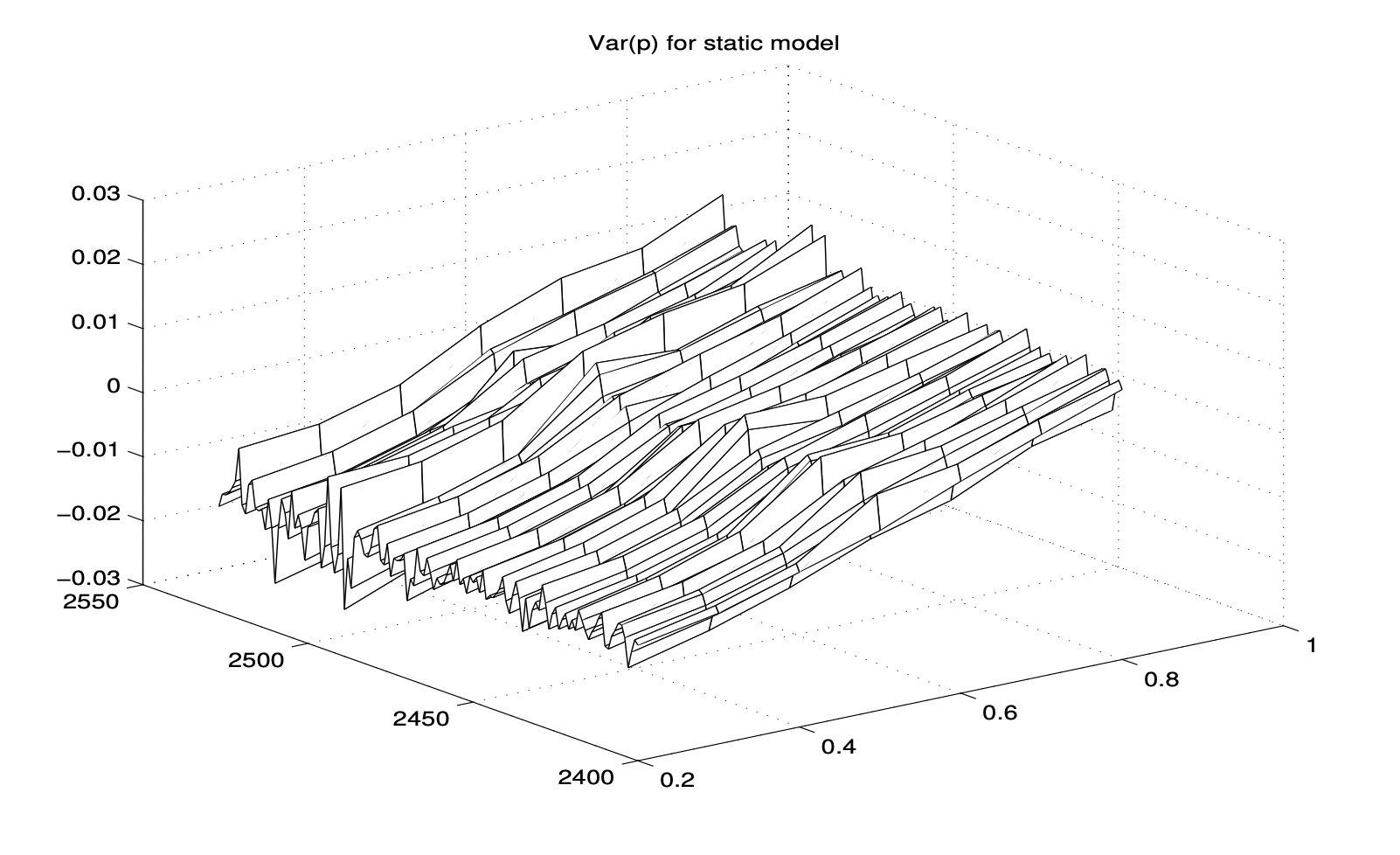} \caption{Non-recursive VaR Surface in time-probability space}
 \end{center}
\end{figure}

\clearpage

\begin{figure}[htbp!] \label{Figure 4}
 \begin{center}
\noindent
 \includegraphics[width=5in,height=3in]{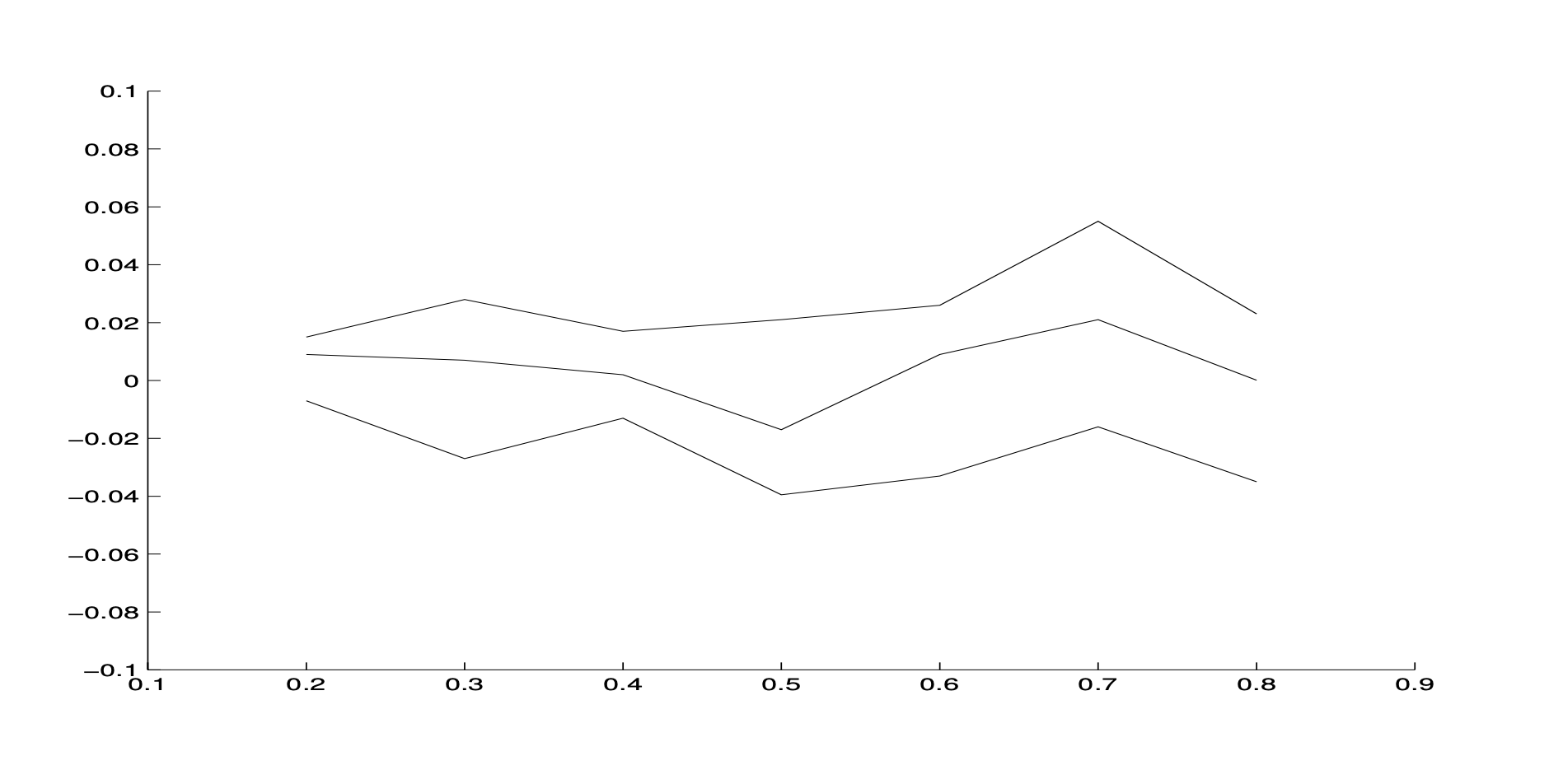} \caption{$\widehat \theta_2(\tau)$ for $\tau \in [.2, .8]$ and
 the $90\%$ confidence intervals.}
 \end{center}
\end{figure}

\begin{figure}[htbp!] \label{Figure 5}
 \begin{center}
\noindent
 \includegraphics[width=5in,height=3in]{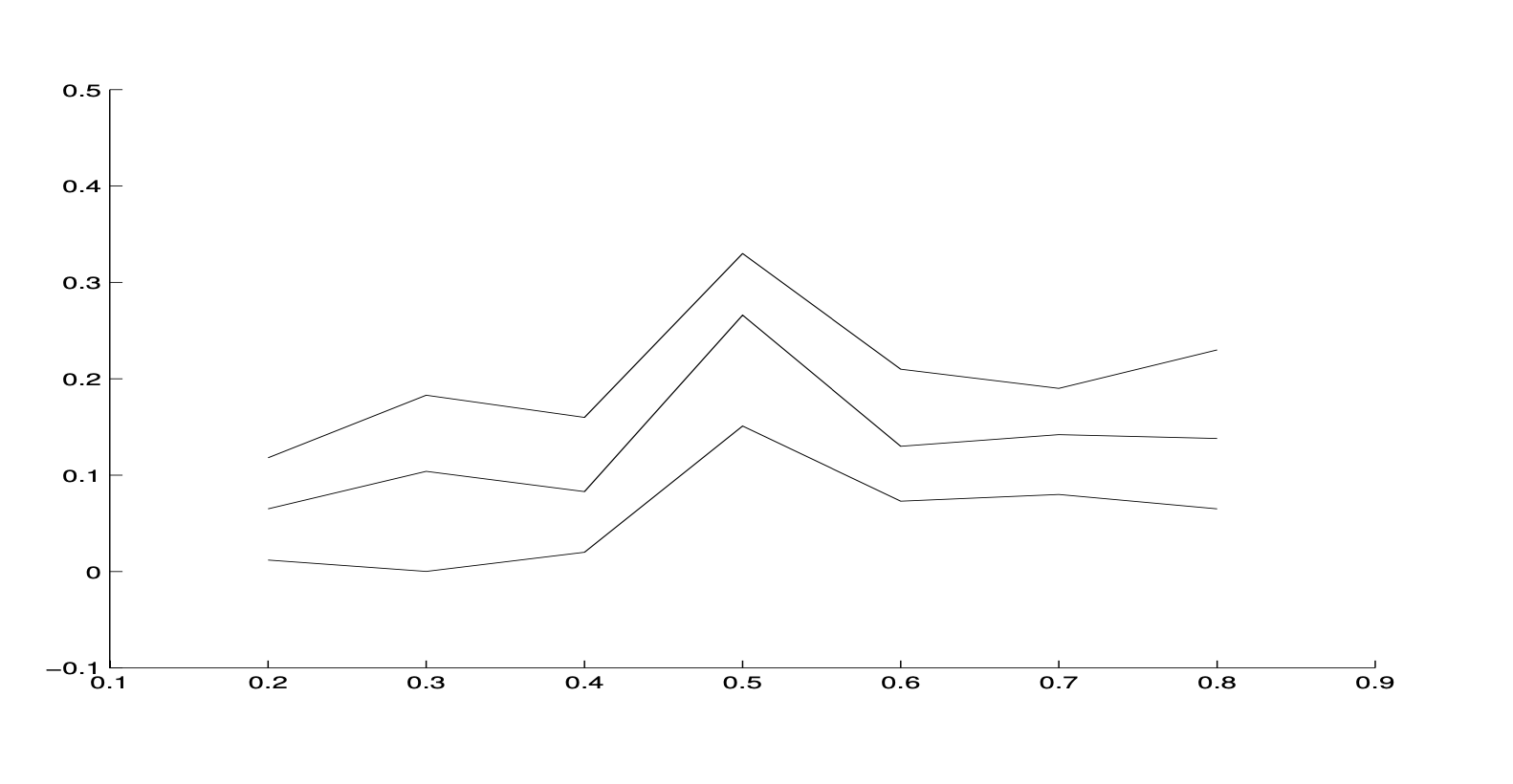}
 \caption{$\widehat \theta_3(\tau)$ for $\tau \in [.2, .8]$ and
 the $90\%$ confidence intervals. }
 \end{center}
\end{figure}
\begin{figure}[htbp!] \label{Figure 6}
 \begin{center}
\noindent
\includegraphics[width=5in,height=3in]{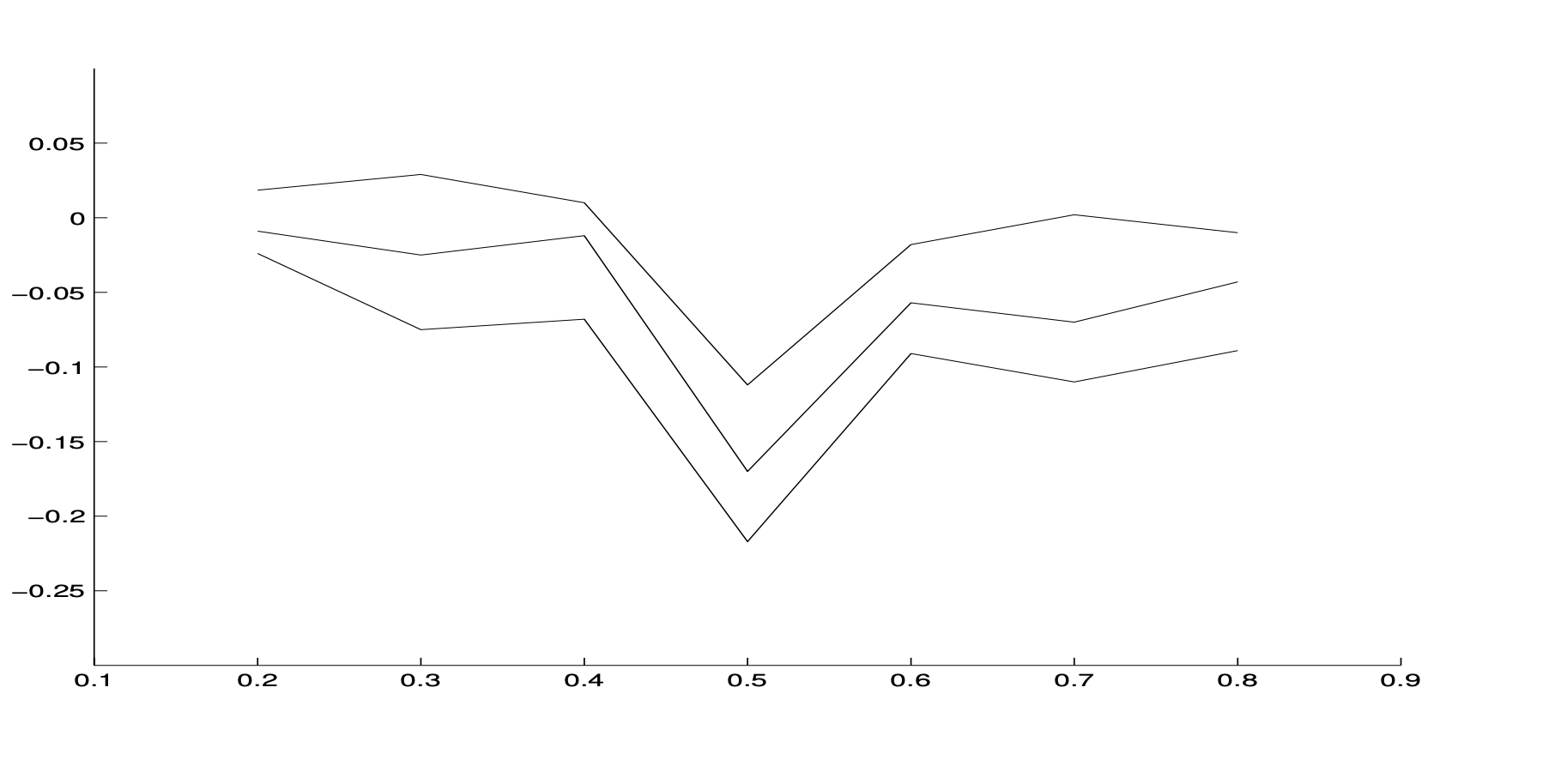} \caption{$\widehat \theta_4(\tau)$ for $\tau \in [.2, .8]$
 and the $90\%$ confidence intervals. }
 \end{center}
\end{figure}
\begin{figure}[htbp!] \label{Figure 7}
 \begin{center}
\noindent
 \includegraphics[width=5in,height=3in]{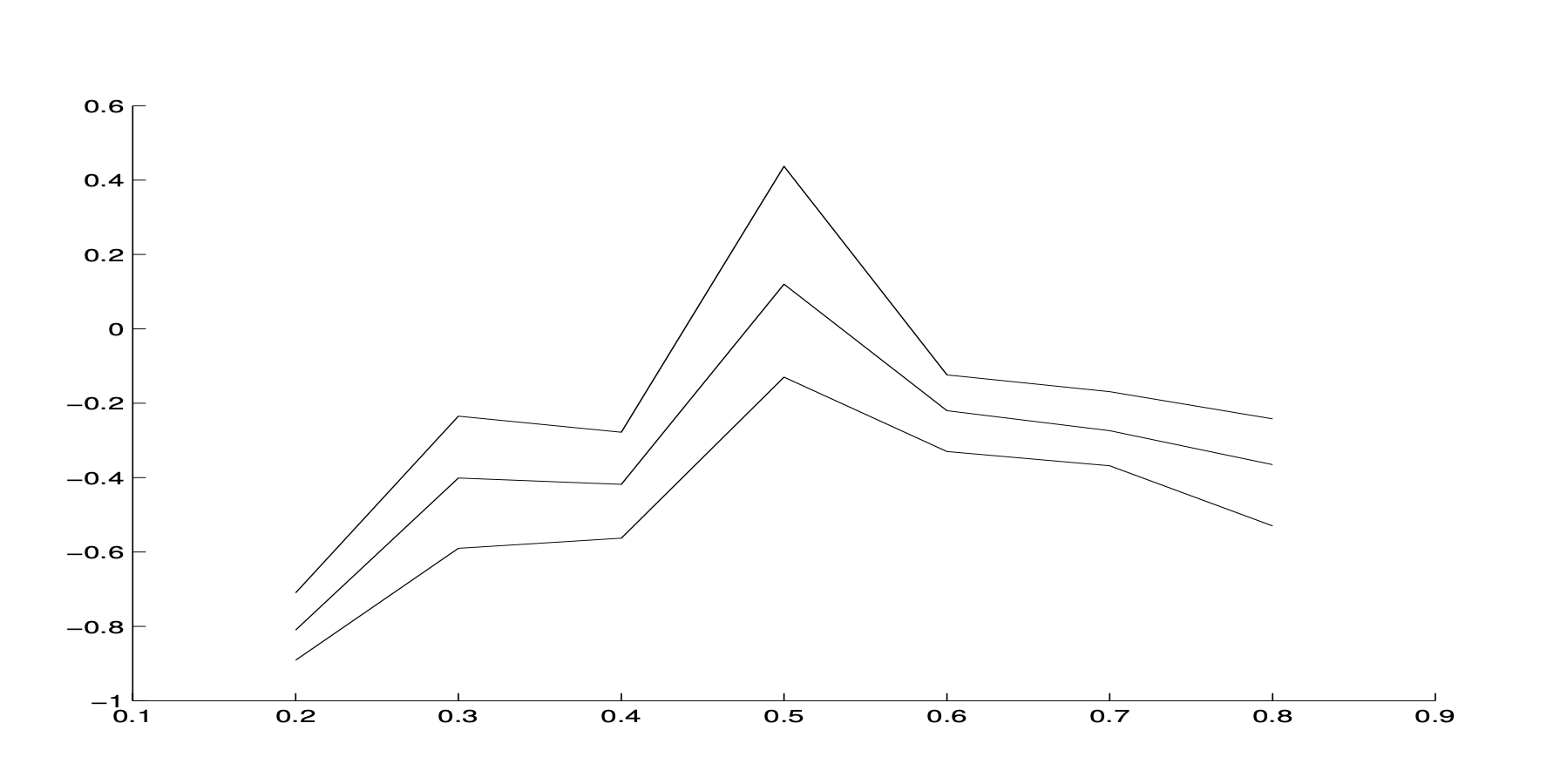} \caption{$\widehat \varrho(\tau)$ for $\tau \in [.2, .8]$
 and the $90\%$ confidence intervals.}
 \end{center}
\end{figure}

\end{document}